\documentclass[10pt]{article}
\usepackage{tgpagella} 

\usepackage{tablefootnote}
\usepackage[dvips]{graphics}
\usepackage[utf8]{inputenc}
\usepackage{amsmath}
\usepackage{amsfonts}
\usepackage{amssymb}
\usepackage{graphicx}
\usepackage{fancyhdr} 
\usepackage{multirow}
\usepackage{wrapfig, framed}
\usepackage{algorithm}
\usepackage{algorithmic}
\usepackage{epsfig,endnotes,color,url,paralist, multirow,float}

\usepackage{subfig}
\usepackage[export]{adjustbox}

\usepackage[left=2cm,right=2cm,top=2cm,bottom=2cm]{geometry}
\usepackage[inline,shortlabels]{enumitem}
\usepackage{mathtools}
\usepackage{mathrsfs} 
\usepackage[table]{xcolor}

\definecolor{lightgray}{gray}{0.9}
\definecolor{LightCyan}{rgb}{0.88,1,1}

\usepackage{lipsum}
\usepackage{comment}
\usepackage{soul}
\usepackage{tikz}
\usepackage{xcolor}
\usepackage[format=plain,
            labelfont={bf,it},
            textfont=it]{caption}
\usepackage{xspace}
\usepackage{hyperref}
\usepackage{booktabs}

\usepackage{cite}
\usepackage{url}

\usepackage[compact]{titlesec}

\newcommand{\ouralg}{\texttt{HICO}\xspace}

\newcommand{\pmtwo}{PM\textsubscript{2.5}\xspace}
\newcommand{\sotwo}{SO\textsubscript{2}\xspace}
\newcommand{\nox}{NO\textsubscript{x}\xspace}
\newcommand{\cotwo}{CO\textsubscript{2}\xspace}
\newcommand{\notwo}{NO\textsubscript{2}\xspace}

\newcommand{\eopt}{\texttt{E-OPT}\xspace}
\newcommand{\copt}{\texttt{C-OPT}\xspace}
\newcommand{\hopt}{\texttt{H-OPT}\xspace}

\newcommand{\hico}{\texttt{HICO}\xspace}

\newcommand{\revise}[1]{{{#1}}}

\makeatletter
\newcommand\mysize{\@setfontsize\notsotiny\@vipt\@viipt}
\makeatother

\begin{document}

\newpage
\pagenumbering{arabic}
\pagestyle{plain}
\setcounter{page}{1}
\setcounter{section}{0}

\begin{center}{{\bf \Large
Health-Informed Computing: Estimating and Addressing the Public Health Impact of Data Centers}}
\end{center}

\vspace{0.5cm}

\begin{table}[!h]
\centering
\begin{tabular}{m{0.16\textwidth} m{0.16\textwidth}m{0.16\textwidth}m{0.16\textwidth} m{0.16\textwidth}}
\centering Yuelin Han\\ \emph{UC Riverside}
&\centering Zhifeng Wu\\ \emph{UC Riverside}
& \centering Pengfei Li\\ \emph{RIT}
& \centering Adam Wierman\\ \emph{Caltech}
& \centering Shaolei Ren\tablefootnote{\;Yuelin Han and Zhifeng Wu
 contributed equally and are listed alphabetically.\\ 
\indent \indent Corresponding authors: Adam Wierman (adamw@caltech.edu) and Shaolei Ren (shaolei@ucr.edu)}\\ \emph{UC Riverside}
\end{tabular}
\end{table}

\begin{center}
    \textbf{Abstract}
    \end{center}
The surging demand for artificial intelligence (AI) has led to a rapid expansion of energy-intensive data centers, contributing to criteria air pollutant emissions
and raising public health concerns that have received comparatively limited attention in sustainability assessments.
 This paper introduces a principled methodology to model air pollutant emissions for data centers and estimate the public health impacts. Our findings show that 
the growing demand for AI and computing technologies is projected to push the total annual public health burden of U.S. data centers up to more than \$20 billion in 2028. 
Although national-level impacts remain modest, data center health costs are
unevenly distributed: in the most affected counties, the estimated per-household health burden
can reach about seven times the national average. 
Next,
we propose a health-informed computing framework that explicitly incorporates public health impacts into data center resource management across space and time, mitigating public health costs while supporting environmental sustainability.
More broadly, we
recommend 
extended energy reporting to include public health impact of data centers and paying attention to all impacted communities.
\vspace{0.3cm}

\section{Introduction}
Artificial intelligence (AI) has significant potential to address major societal challenges, including air quality, public health,
disease prevention and healthcare optimization. At the same time, the rapid growth of generative AI, particularly large language models (LLMs), has sharply increased computational demand and accelerated the expansion of energy-intensive data centers. According to the recent U.S. data center energy report
\cite{DoE_DataCenter_EnergyReport_US_2024}, AI workloads are projected to proliferate and, together with other growing computing demands, could raise U.S. data center electricity consumption to 6.7--12.0\% of the national total by 2028, up from 4.4\% in 2023.

This surge in electricity demand places growing stress on power grids \cite{DoE_AI_DataCenter_EnergyDemand_Recommendation_2024} and intensifies environmental impacts through increased carbon emissions \cite{Carbon_SustainbleAI_CaroleWu_MLSys_2022_wu2022sustainable} and water consumption \cite{Shaolei_Water_AI_Thirsty_CACM}. While mitigation strategies such as grid-interactive data centers, energy-efficient hardware and software, and carbon- and water-aware computing have been explored \cite{DataCenter_FlexibilityInitiative_EPRI_WhitePaper_2024,Shaolei_Water_AI_Thirsty_CACM}, these efforts have yet to incorporate another critical dimension: Public Health.

\textbf{The underexamined public health impact of data centers.}
The reliance on fossil fuel-based electricity
and usage of on-site (typically diesel) generators
to operate data centers
contribute to air quality concerns and public health costs through the emission of criteria air pollutants. 
These pollutants include fine particulate matter (\pmtwo), sulfur dioxide (\sotwo), and nitrogen dioxide (\notwo). 

Exposure to criteria air pollutants, especially \pmtwo, is causally linked to premature mortality, asthma, cardiovascular disease, 
and other health effects, with
adverse impacts observed even at \pmtwo concentrations below national air quality standards (``no-threshold'') \cite{EPA_PM25_NoThreshold_Ozone_HealthBenefits_TSD_2024}.  
Further, criteria air pollutants are not confined to the immediate vicinity of their emission sources; they can travel hundreds of miles through atmospheric dispersion (i.e., cross-state air pollution) \cite{Health_COBRA_EPA_Mannual}.

Globally, ambient air pollution caused 4.2 million premature deaths in 2019 and remains one of the leading risk factors for disease burden across all socio-demographic groups \cite{Health_AirPollution_Ambient_4dot2_million_PrematureDeath_2019_WHO_Website,Health_ParticulateMatterLeadingFactor_GlobalBurdenDisease2021_IMHE_2024}.  
While the U.S. has generally better air quality than many other countries, 
4 in 10 people in the U.S. still live with unhealthy levels of air pollution,
according to the ``State of the Air 2024'' report published
by the American Lung Association
\cite{Health_4in10_Americans_UnhealthAir_US_AmericanLungAssociation_report_2024}.
In 2019 (the latest year of data provided by the World
Health Organization, or WHO, as of November 2024), an estimate of 93,886 deaths in the U.S. were attributed to ambient air pollution \cite{Health_AirPollution_Ambient_AttributableDeath_Global_WHO_Website}.

Electricity generation, along with transportation and industrial activities, is an important contributor to ambient air pollution with significant public health consequences \cite{EPA_PowerPlant_Health_LeadingSource_Website}. For example,  a recent study \cite{Health_CoalPowerPlants_US_Harvard_UTAustin_Science_2023_doi:10.1126/science.adf4915} found that   460,000 \emph{excess} deaths were attributed to \pmtwo emissions from coal-fired power plants between 1999 and 2020 in the U.S. alone. As emphasized by the U.S. Environmental Protection Agency (EPA), despite decades of progress, ``fossil fuel-based power plants remain a leading source of air, water, and land pollution that affects communities nationwide'' \cite{EPA_PowerPlant_Health_LeadingSource_Website}.

  Looking forward, the electricity sector is generally expected to become cleaner over time, but the pace, scale, and regional distribution of this transition may remain limited \cite{EIA_EnergyOutlook_2026_website}. For example,
  the U.S. Energy Information Administration's 2026 Annual Energy Outlook projects that coal consumption by the electricity sector in 2050 remains approximately 40\% of its 2025 level in the ``Alternative Electricity'' case, which assumes that EPA's April 2024 power-plant \cotwo rule is not in place~\cite{EIA_EnergyOutlook_2026_website}. 
 At the global scale, 
electricity generation has remained heavily dependent on coal and other fossil fuels, underscoring the persistent challenge of fully powering data centers with pollutant-free energy 
\cite{ElectricityMix_FossilFuel_OurWorldinData_2024}. 
Moreover, the growing energy demand of data centers is already delaying the decommissioning of coal-fired power plants and driving the expansion of fossil-fuel power plants in some regions  
\cite{DoE_AI_DataCenter_EnergyDemand_Recommendation_2024,Energy_IntegratedResourcePlan_Dominion_Virginia_2024}.

Addressing air-pollution-related health impacts requires coordinated efforts across sectors, along with mitigation strategies tailored to each sector \cite{Health_COBRA_EPA_Website}. While health impacts of air pollution from sectors such as transportation have been widely studied \cite{Health_ElectricVehicle_PowerPlant_Toronto_PNAS_2024_schmitt2024health}, the public health impacts of data centers have received comparatively less attention
and often remained absent from infrastructure risk assessments and sustainability reports. Without effective mitigation strategies, the health impacts, including hospital admissions, asthma symptoms, and mortality, are likely to grow with rising data center demand.

\textbf{Estimating and addressing public health impacts.} To address the gap in the literature,
we introduce a novel methodology to estimate the hidden public health impacts of data centers. 
Specifically, focusing on the contiguous United States,
we use the EPA's COBRA model \cite{Health_COBRA_EPA_Website} to estimate pollutant dispersion and resulting health outcomes attributed to data centers
associated with their backup generation (Scope 1) and electricity usage (Scope 2).
Our estimates show that U.S. data centers could contribute to various health outcomes including approximately 600,000 asthma symptom cases and 1,300 deaths in 2028
under the high-growth scenario, with total public health costs exceeding \$20 billion.
This corresponds to an increase of 213\% relative to the 2023 level,
compared with a projected 17\% increase in U.S. stationary fuel-combustion-related health
costs over the same period. 
Although national-level impacts remain modest, data center health costs vary substantially across counties:
the highest county-level per-household
health cost is about seven times the national average and approximately 200 times the
lowest county-level value,
warranting
closer attention.

To help mitigate the growing public health burden, we propose \emph{Health-Informed Computing} (\ouralg), which leverages data center flexibility and explicitly incorporates public health costs into siting and resource management decisions. Using spatial load shifting as a case study, we show that \ouralg can
substantially reduce health costs while complementing the broader
sustainability goals of existing carbon-aware computing.
\ouralg also aligns well with demand-side energy innovations aimed at improving public health \cite{Health_Benefit_kWh_EPA_Website}.

Finally,
we provide broader recommendations to address the increasing public health impact of data centers, including extended energy reporting to include public health assessment 
and paying attention to all impacted communities.

\textbf{Disclaimer.} \emph{The results presented in this paper are not intended to encourage or discourage the construction of data centers,  nor should they be used to support or oppose any specific project, which requires more detailed and context-specific evaluation.
We do not take a position on decisions related to any specific data centers or the use of AI.
Instead, our goal is to provide a quantitative assessment of the potential public health impacts of the data center industry and to develop health-informed computing as a mitigation strategy that can help reduce these impacts while supporting sustainable growth. Throughout this paper, terms such as ``health costs,'' ``health impacts,'' and ``health burden'' refer to population-level estimates produced using the U.S. EPA's screening model COBRA (Desktop v5.1), rather than observed health outcomes or individually attributable effects.}

\section{Methodology of Estimating Public Health Impacts of Data Centers}\label{sec:background_scope}

\begin{figure}[!t]
    \includegraphics[width=1\textwidth, trim={0cm 0.7cm 0 0cm},clip]{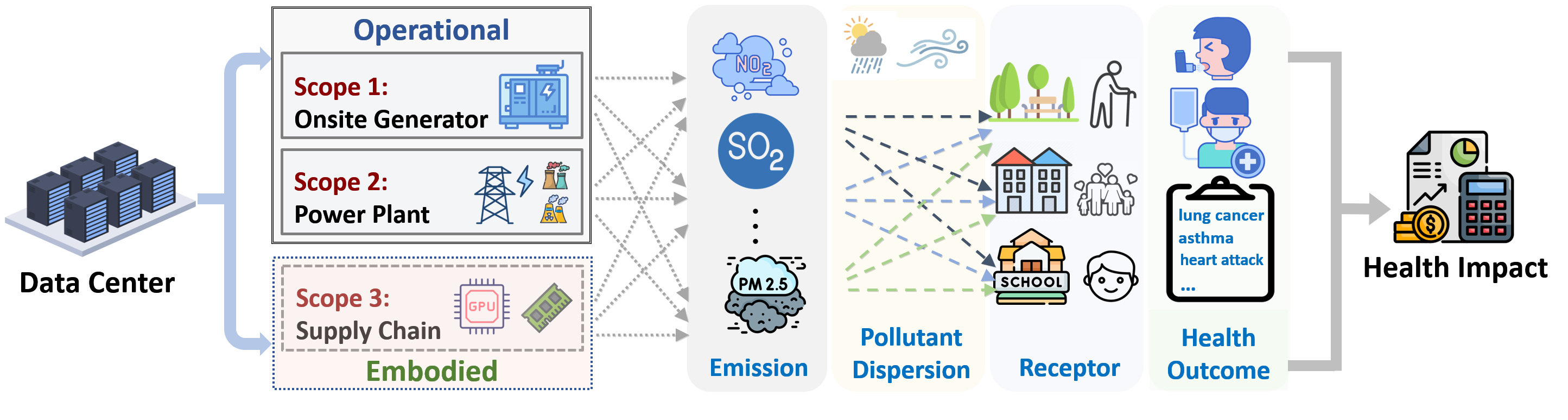}   
    \vspace{-0.5cm}
        \caption{The overview of data centers' contribution to air pollutants and public health impacts. Scope-1 and scope-2 impacts occur during the operation of data centers (``operational''), whereas scope-3 impacts arise from activities across the supply chain (``embodied'').}
        \label{fig:health_overview_diagram}
\end{figure}

This section presents our methodology of estimating
data centers' contribution to criteria air pollutants 
and public health impacts
throughout its lifecycle across three scopes (Fig.~\ref{fig:health_overview_diagram}). 
The scoping definition
in this paper parallels the well-established greenhouse gas protocol \cite{Carbon_Scope_Definition_Website}.

\subsection{Background on Air Pollutants}

Criteria air pollutants, including \pmtwo, \sotwo and \notwo, 
are a group of airborne contaminants that are emitted from various sources
such as industrial activities and vehicle emissions.
The direct emission of \pmtwo is called
 primary \pmtwo, while precursor pollutants such as \sotwo, \nox, and VOCs,
 can form secondary \pmtwo and/or ozone. 
These air pollutants can travel a long distance (a.k.a. cross-state air pollution),
posing direct and significant risks to public health over large areas,
particularly for vulnerable populations including the elderly and individuals with respiratory conditions  \cite{Health_COBRA_EPA_Mannual}.

Under the Clean Air Act, the U.S. EPA is authorized to regulate the emission levels of criteria air pollutants, 
reducing concentrations
to comply with the National Ambient Air Quality Standards (NAAQS).
For example, the NAAQS primary standards
set the annual average \pmtwo concentration at 9$\mu g/m^3$
and the 98-th percentile of 1-hour daily maximum \notwo 
concentration at 100 parts per billion by volume, both counted over three years \cite{EPA_AirQuality_Table_Website}.
In addition, state and local governments may set additional regulations on criteria air pollutants to strengthen or reinforce national standards \cite{California_AirResourcesBoard_website}.  

The U.S. EPA treats \pmtwo as a ``no-threshold'' pollutant, meaning that adverse public health impacts can occur even at concentrations below national or regional air quality standards
\cite{EPA_PM25_NoThreshold_Ozone_HealthBenefits_TSD_2024}.
Thus, regulatory compliance does not necessarily imply the absence of health risk.
For example, 
the U.S. standard for the annual average limit of \pmtwo 
is still higher than the WHO's recommended level of 5 $\mu g/m^3$~\cite{Health_WHO_AirQuality_Guideline_PM25_2021,EPA_AirQuality_Table_Website}. 

While \cotwo is broadly classified
by the U.S. EPA as an air pollutant following the U.S. Supreme Court ruling in 2007
\cite{EPA_Carbon_AirPollutant_SupremeCourtRuling_Massachusetts_LibraryOfCongress_2007}, it 
often does not cause the same immediate health impacts as criteria pollutants. 
Thus, 
we use
``air pollutants'' to solely refer to criteria air pollutants wherever applicable.

\subsection{Modeling Tool}

We focus on the United States,
one of the world's largest data center markets. 
The U.S. EPA provides  COBRA, a convenient tool
for assessing public health impacts of air pollutants
in contiguous U.S. (simply referred to as the U.S. in this paper)
\cite{Health_COBRA_EPA_Website}.  
By taking the amount
of emissions at the source as the input,
 COBRA performs simplified air dispersion modeling
(including both primarily emitted \pmtwo and secondarily formed
\pmtwo and ozone)
with various concentration-response functions \cite{Health_COBRA_EPA_Mannual}, 
 offering a quantitative 
analysis of county-level public health impacts.\footnote{Independent cities considered
county-equivalents for census purposes are also referred to as ``counties'' in COBRA.}

Specifically, COBRA uses a simplified source-receptor (S-R) matrix to model
air dispersion, i.e., the movement
of emitted pollutants in the atmosphere. It then estimates a range of health outcomes based
on epidemiological evidence, including mortality, heart attacks, and asthma
symptoms \cite{Health_COBRA_EPA_Website}. For example, mortality is estimated
using a log-linear concentration-response function for \pmtwo, with $\beta=0.011330$ and $\beta=0.006390$ for the high and low estimates, respectively
\cite{Health_COBRA_EPA_Mannual}. Finally, COBRA assigns an economic value to
each health outcome and aggregates these values to estimate the overall public
health burden associated with a pollutant-emitting activity. Further details of
COBRA are provided in the supporting manual \cite{Health_COBRA_EPA_Mannual}.

Although COBRA is a screening model and cannot replace more sophisticated air quality models for site-level regulatory decisions, its estimates have been shown to be
reasonably consistent with those from advanced models \cite{Health_COBRA_EPA_Mannual}, making it well
suited for regional public health impact assessment. COBRA has also been widely used in the literature to study
the health impacts of various sectors, including transportation and electricity generation \cite{EPA_Papers_Cite_COBRA_Website}.

We use COBRA Desktop v5.1 for our analysis. 
All monetary values are reported in 2023 U.S. dollars,
using COBRA's default discount rate of 2\% \cite{Health_COBRA_EPA_Website}.
We report COBRA results in the format of ``mid (low, high),'' where the three
values correspond to the midrange, low, and high estimates,
respectively. When presenting a single value or ratio, we use the midrange estimate by default. More details are available in Appendix~\ref{sec:methodology_details}.

\subsection{Data Centers' Contribution to Air Pollutant Emissions}

As illustrated in Fig.~\ref{fig:health_overview_diagram}, data centers contribute to criteria air pollutant emissions through three scopes, which serve as the key inputs to COBRA for estimating their public health impacts.

\subsubsection{Scope~1: Onsite Generator}\label{sec:background_scope_1}
Data centers are mission-critical facilities with high availability requirements and need reliable backup power sources \cite{Google_SustainabilityReport_2024,Facebook_SustainabilityReport_2024}.
 Due to the limited experience
with cleaner backup alternatives at scale \cite{DoE_AI_DataCenter_EnergyDemand_Recommendation_2024}, 
many data centers, including newly built facilities, continue to depend on onsite diesel generators for backup power \cite{DoE_AI_DataCenter_EnergyDemand_Recommendation_2024,Virginia_AirPermitsDataCenter_Website}. Nonetheless,
diesel generators are known to emit significant amounts of air pollutants during operation \cite{EPA_StationaryEngines_BackupGenerators_Rule_Website}.

In Northern Virginia, which has the world's largest concentration of data centers, most of the diesel backup generators, including many recently installed units, are classified as Tier~2 \cite{Virginia_AirPermitsDataCenter_Website}.
Tier~4 generators use more advanced exhaust aftertreatment and therefore have lower emission rates than Tier~2 units, but they often add cost, complexity, and operational constraints \cite{EPA_Diesel_Nonroad_HeavyEquipment_Regulations_Tier4_Website}. 
 As of the end of 2024, the total permitted annual emission limits for these on-site generators were approximately 13,000 U.S. short tons of \nox, 1,400 tons of VOCs, 50 tons of \sotwo, and 600 tons of \pmtwo.

While backup generators typically do not operate for extended periods, regular testing and maintenance are necessary for reliability. In 2023, backup generators at Virginia data centers emitted about 7\% of their permitted amounts, primarily for maintenance (often 10 to 20 hours per year) \cite{Virginia_AirPermitsDataCenter_Report_JLARC_2024}. Similarly, historical emissions at some data centers in Quincy, Washington reached 3\%--12\% of permitted levels \cite{Washington_AirQualityDataCenter_Report_2020}.
More recently, the U.S. EPA clarified that data center backup generators may be operated for up to 50 hours per year for non-emergency demand response or reliability-related purposes, provided that specific regulatory conditions are met and the operation complies with applicable air permits \cite{EPA_50hour_Rule_Clarification_Guidance_Webpage}.

According to a recent Virginia state report \cite{Virginia_AirPermitsDataCenter_Report_JLARC_2024}, during an extended outage, an affected data center could potentially reach its annual emissions limits ``within a few days.'' More broadly, when many generators in a region operate simultaneously during grid stress, they can produce short-term spikes in local air pollutants. For example, in an upper-bound case where data center backup generators in Northern Virginia emit air pollutants at their maximum permitted levels during a prolonged regional grid outage,
 their \nox emissions could amount to more than half of the region's annual \nox emissions from all sources \cite{Virginia_AirPermitsDataCenter_Report_JLARC_2024}.

To account for scope-1 emissions, we consider a reference case in which actual emissions are 10\% of permitted levels, informed by both historical reports and potential future demand-response use. If the actual percentage is $x\%$, our estimate scales approximately by $\frac{x}{10}$. We derive an average annual emission rate in tons/MWh from Virginia data centers (representing approximately 20\% of U.S. data center electricity use in 2023) \cite{Virginia_AirPermitsDataCenter_Website,DataCenter_Energy_EPRI_AI_9Percent_US_2030_WhitePaper_2024} and apply it to data center electricity consumption in other states,
because scope-1 emissions data for data centers in other states are often less readily available. The effect of this approach on our overall results is expected to be limited because scope-2 health costs are significantly larger in our reference case.

\subsubsection{Scope~2: Power Plant}\label{sec:background_scope_2}

Data centers' scope-2 public health impacts depend on where the power plants supplying their electricity are located.
For national-scale analysis, we use the average attribution method unless otherwise stated, which
is also widely adopted by technology companies for carbon emission accounting \cite{Google_SustainabilityReport_2024}.
We focus on location-based accounting without considering market-based offsetting mechanisms, which may be less effective in mitigating health impacts from location-dependent emission sources.

The U.S. electricity grid is
divided into 14 regions following the AVoided Emissions and geneRation Tool (AVERT v4.3) provided by the U.S. EPA \cite{Health_AVERT_Emission_Marginal_EPA_Website}.
We first calculate
the total data center electricity consumption $e_{DC}$ and the overall
electricity consumption (including non-data center loads) $e_{Total}$ within
each electricity region. 
We use the state-level data center electricity consumption distribution for 2023 provided
by EPRI \cite{DataCenter_Energy_EPRI_AI_9Percent_US_2030_WhitePaper_2024}, scale it by the U.S. total data center electricity consumption
in 2023 and for the 2028 projection based on data provided
by \cite{DoE_DataCenter_EnergyReport_US_2024},
and then distribute state-level electricity consumption to relevant electricity regions following the state-to-region electricity apportionment used by AVERT.

Next, we calculate the percentage $x\%=\frac{e_{DC}}{e_{Total}}$ of the data center electricity consumption with respect to the total electricity consumption for each region. The relationship between
the health impact and emission reduction in COBRA is approximately linear.
Thus, 
we apply a reduction by $x\%$
to the baseline emissions
of all the power plants within the respective electricity region in COBRA
and estimate the corresponding county-level health impacts. 

Although co-location with on-site, often temporary and gas-fired, power plants is an emerging option for mitigating grid capacity constraints, it remains limited today. Thus, our average attribution of data center electricity use to regional off-site power plants remains a reasonable approximation for our national-scale analysis. Moreover,
depending on the technology, pollutant control, and operational settings,
 on-site plants
may also have emission rates comparable to regional grids, raising potential public health concerns \cite{Health_Virginia_DEQ_Vantage_AirPollutant_Onsite_Gas_Report_Rebuttal_2026}.

\subsubsection{Scope 3: Supply Chain}\label{sec:background_scope_3}

Scope-3 air pollutants include embodied supply-chain emissions, such as those associated with manufacturing GPUs and servers. Because these emissions are harder to quantify due to limited public data, we focus on scope-1 and scope-2 health impacts in this paper, which we collectively refer to as ``operational'' impacts.
A detailed assessment of the health impacts associated with a specific U.S. semiconductor manufacturing
facility is provided in Appendix~\ref{sec:semiconductor_calculation}.

\section{Public Health Impact of U.S. Data Centers}

We present our public health impact analysis for U.S. data centers, beginning with the scope-1 impacts of Virginia data centers, followed by the scope-1 and scope-2 impacts of U.S. data centers in 2023 and projections for 2028.
\revise{We focus on 2023 and 2028, because 
COBRA provides county-level population, health incidence, valuation, and baseline emissions data for both years \cite{Health_COBRA_EPA_Website}
and they are also used as the reference years in the recent U.S. data center energy report \cite{DoE_DataCenter_EnergyReport_US_2024}.}

Importantly, the 2023 and 2028 baseline emissions used by COBRA are based on the U.S. EPA's Emissions Modeling Platform 2016v1 and account for federal and state regulations as of May 2018 \cite{Health_COBRA_EPA_Website}. However, regulatory and emissions changes since then may introduce substantial uncertainty into the underlying emission projections, warranting future re-analysis where applicable. Accordingly,
our COBRA-based estimates should not be interpreted as assigning responsibility for any specific individual health outcome, establishing site-specific impacts, or implying that data centers are a dominant contributor to national air-pollution-related health burdens. Rather, they represent a modeled, first-order approximation of how data center-related emissions may contribute to population-level public health impacts, with the goal of informing potential mitigation strategies.

\subsection{Scope-1 Health Impact of Virginia Data Centers}

The high emission rate from onsite backup generators, mostly powered by diesel fuel, could pose health risks,
especially
in regions with a concentration of large data centers.
To illustrate this point, we consider the data centers' onsite generators in Virginia.
 Assuming that the actual emissions are 10\% of the permitted level (as of December, 2024) as a reference case,
our estimates show that the backup generators could contribute to approximately 14,000 asthma symptom cases and 13--19 deaths each year, among other health outcomes, resulting in an annual public health impact of \$220--300 million throughout the U.S.
In the hypothetical worst-case scenario where data centers' onsite generators in Virginia emit air pollutants at their full annual permitted levels (e.g., 
during a prolonged regional grid outage), the estimated annual public health cost would be \$2.2--3.0 billion.

\begin{figure}
    \centering
    \subfloat[Per-household health cost]{\includegraphics[height=0.235\textwidth,valign=b,trim={18cm 6cm 0 0cm},clip]{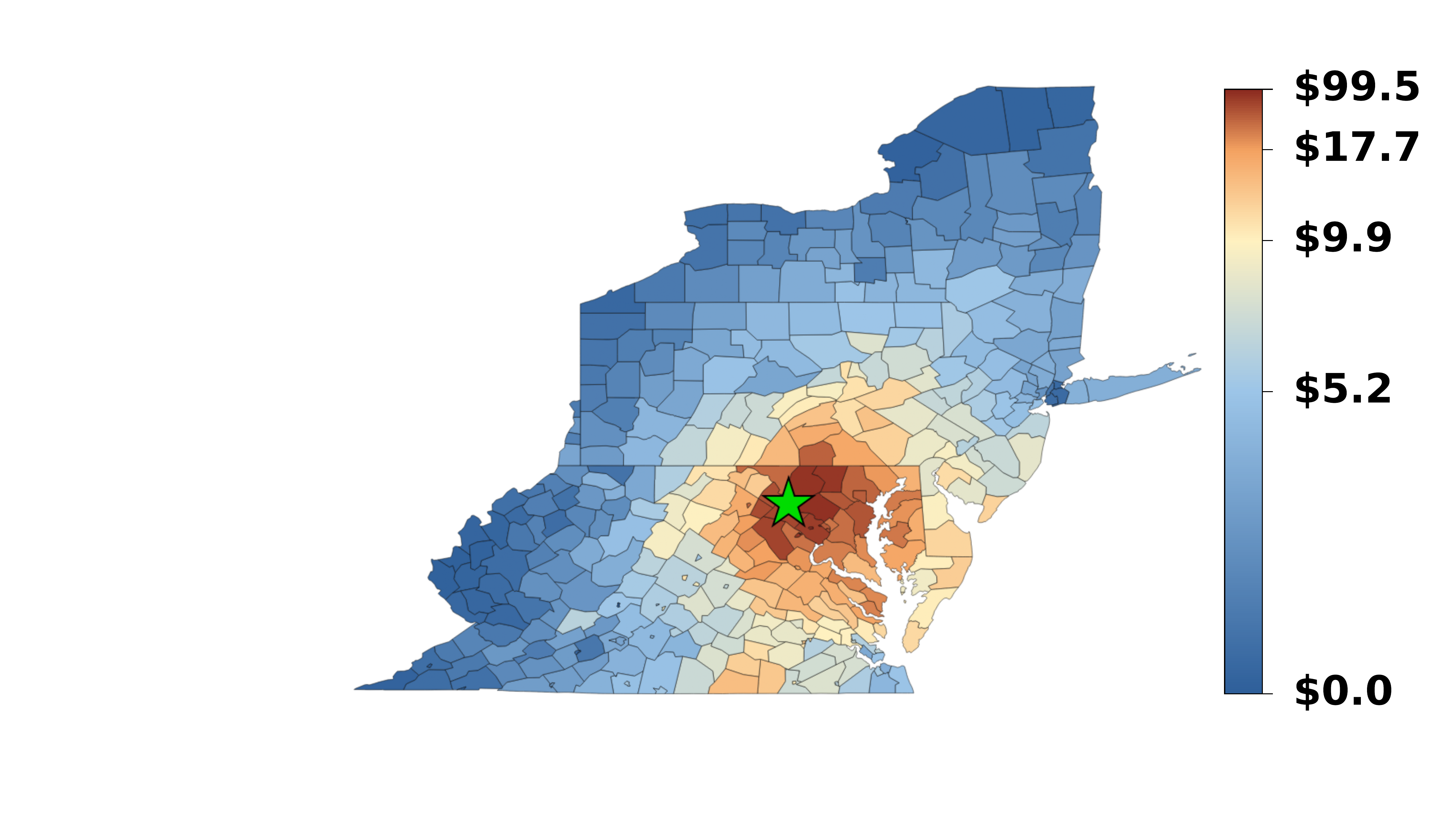}\label{fig:generator_VA_health_2023/map_generator_health_per_household}}
    \subfloat[Top-10 counties]{
        \adjustbox{height=0.11\textwidth,valign=b}{
        \tiny
        \begin{tabular}{c|c|c} 
        \toprule
        \textbf{State} & \textbf{County} & \begin{tabular}[c]{@{}c@{}}\textbf{Per-Household}\\\textbf{Health Cost (\$)}\end{tabular} \\ 
        \hline
        VA & Fairfax City & \textbf{99.5} (85.8, 113.1) \\ \hline
        VA & Falls Church & \textbf{58.3} (46.6, 69.9) \\ \hline
        MD & Montgomery & \textbf{46.9} (40.9, 53.0) \\ \hline
        MD & Frederick & \textbf{44.1} (37.6, 50.6) \\ \hline
        MD & Carroll & \textbf{41.0} (35.0, 46.9) \\ \hline
        VA & Manassas City & \textbf{37.2} (32.1, 42.3) \\ \hline
        VA & Fairfax & \textbf{36.4} (31.9, 40.9) \\ \hline
        VA & Manassas Park & \textbf{35.4} (29.9, 40.8) \\ \hline
        VA & Fauquier & \textbf{29.8} (25.3, 34.4) \\ \hline
        VA & Loudoun & \textbf{29.6} (25.9, 33.4) \\ 
        \bottomrule
        \end{tabular}}}
        \caption{County-level scope-1 per-household health costs from data center backup generators in Virginia under the reference case, assuming emissions equal 10\% of total permitted emissions as of December 2024. (a) 
    Choropleth map. Loudoun County, Virginia, which has the largest concentration of data centers, is marked in a green star. The legend shows percentile anchors at the 0th, 50th, 75th, 90th, and 100th percentiles, with colors interpolated between anchors. (b) Top-10 counties by per-household health cost.}
    \label{fig:backup_generator_northern_VA_per_household}
\end{figure}

We show the county-level per-household health cost and the top 10 counties in Figure~\ref{fig:backup_generator_northern_VA_per_household}. The results suggest that elevated health impacts can occur not only in host communities, but also in downwind communities, including those far from the data centers, due to long-range pollutant transport. The choropleth map of county-level total health costs is provided in Appendix~\ref{sec:backup_generator_northern_virginia}.

\subsection{Health Impact of U.S. Data Centers in 2023}

\begin{figure}[!t]
\centering
\subfloat[Total health cost]{
    \includegraphics[height=0.2\textwidth,valign=b]{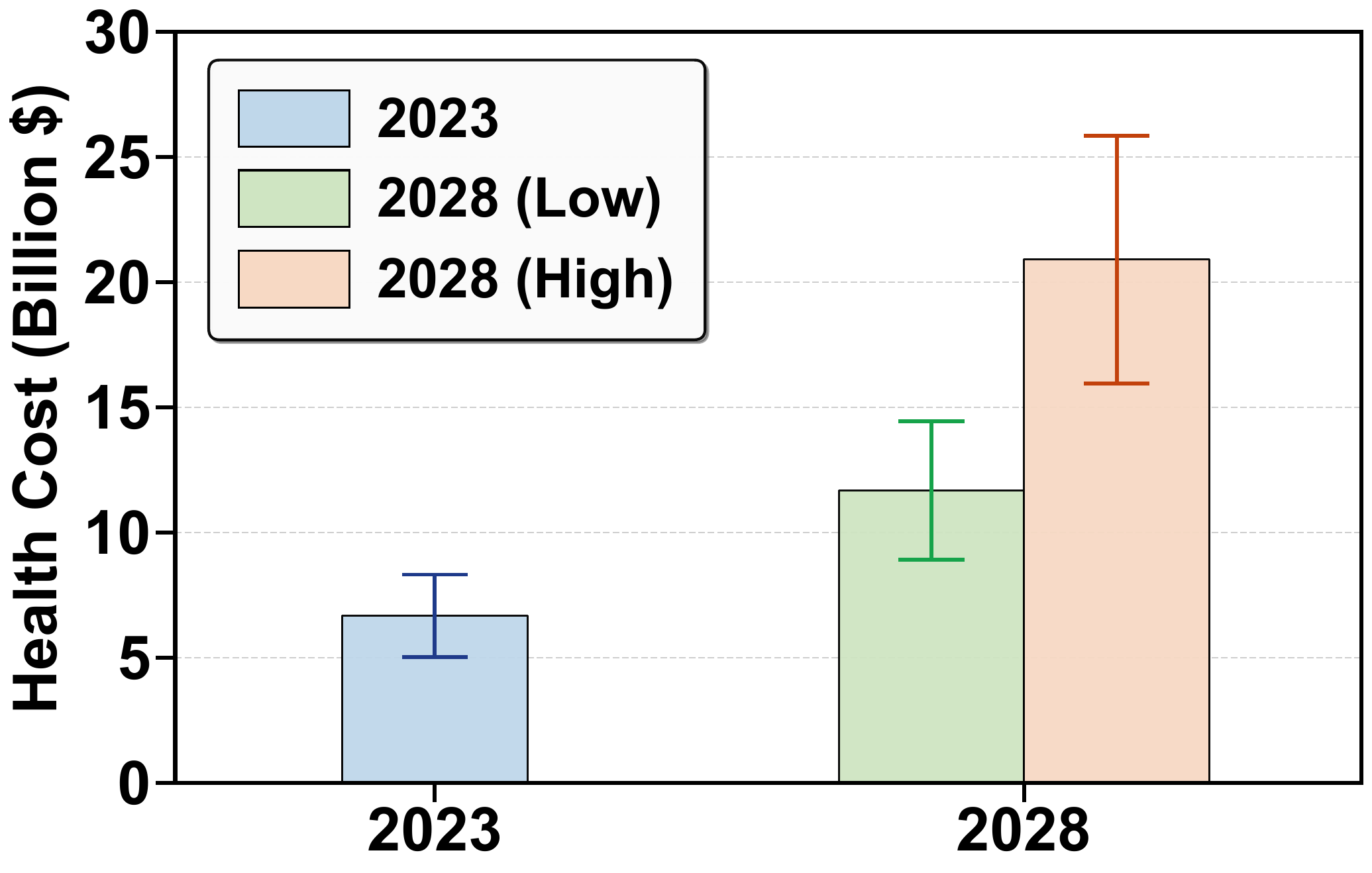}
     \label{fig:total_health_cost_US_2023_2028}
    }
    \subfloat[Per-household health cost]{
    \includegraphics[height=0.2\textwidth,valign=b]{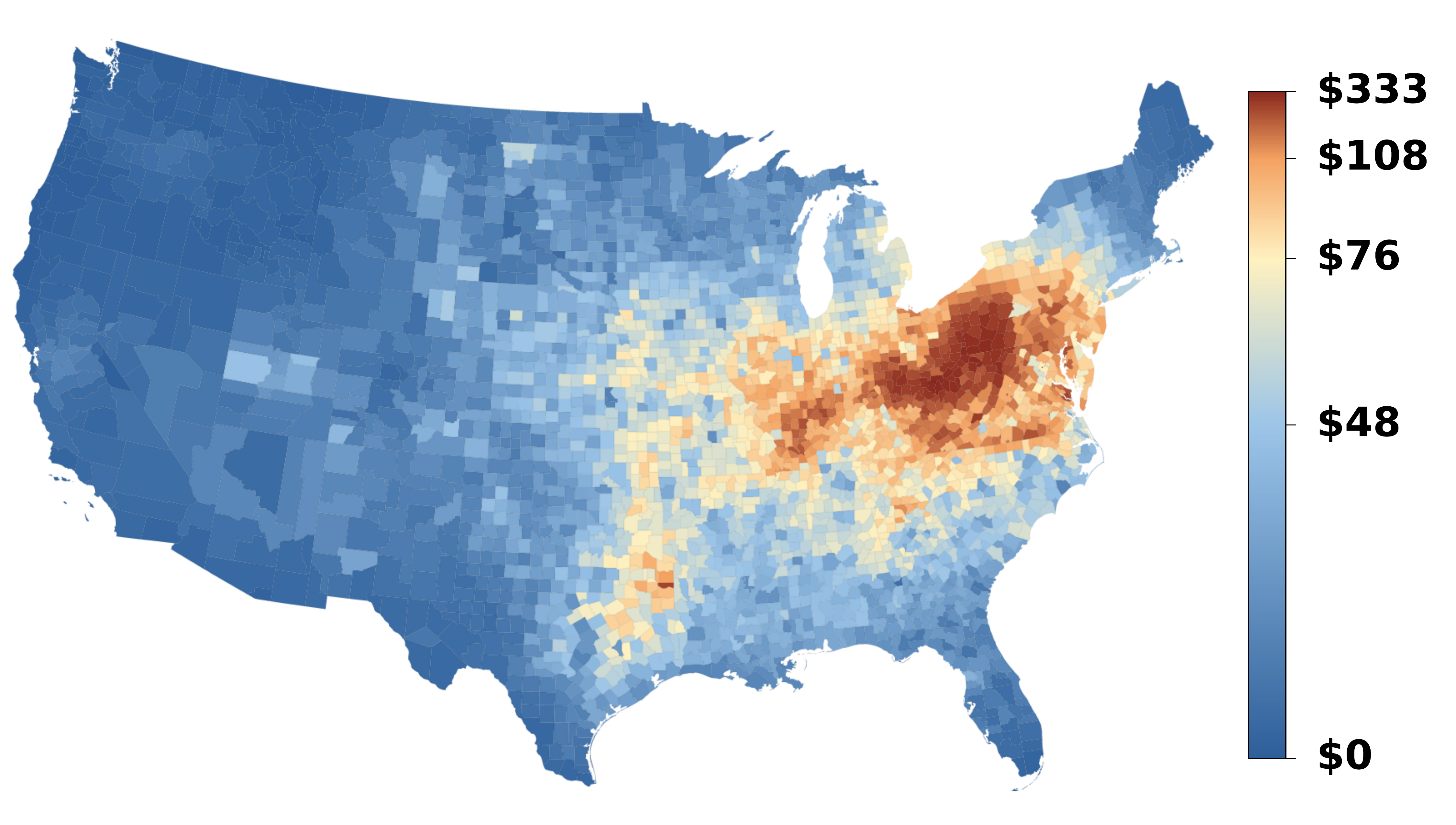} 
    \label{fig:county_percapita_dc_2023}
    }
    \subfloat[Top-10 counties]{
        \adjustbox{height=0.11\textwidth,valign=b}{
        \centering
        \tiny
        \begin{tabular}{c|c|c} 
        \toprule
        \textbf{State} & \textbf{County} & \begin{tabular}[c]{@{}c@{}}
        \textbf{Per-Household} \\ \textbf{Health Cost} (\$)\end{tabular} \\ 
        \hline
        WV & Marion & \textbf{332.8} (266.6, 399.0) \\ 
        \hline
        WV & Mason & \textbf{322.6} (254.2, 391.1) \\ 
        \hline
        OH & Meigs & \textbf{317.5} (237.6, 397.4) \\ 
        \hline
        OH & Gallia & \textbf{312.9} (234.0, 391.8) \\ 
        \hline
        WV & Marshall & \textbf{307.2} (236.5, 377.8) \\ 
        \hline
        WV & Taylor & \textbf{289.8} (234.4, 345.3) \\ 
        \hline
        PA & Fayette & \textbf{279.4} (220.5, 338.3) \\ 
        \hline
        PA & Greene & \textbf{267.4} (218.4, 316.4) \\ 
        \hline
        WV & Brooke & \textbf{258.6} (195.7, 321.6) \\ 
        \hline
        WV & Jackson & \textbf{245.5} (198.2, 292.7) \\ 
        \bottomrule
        \end{tabular}
        }}
        \vspace{-0.1cm}
        \caption{Scope-1 and scope-2 public health costs of U.S. data centers.
        (a) Total health costs in 2023 and 2028. ``High'' and ``Low'' denote the high- and low-growth scenarios  in \cite{DoE_DataCenter_EnergyReport_US_2024}, respectively. 
        (b) 
        County-level per-household health cost distribution in 2023. The legend shows percentile anchors at the 0th, 50th, 75th, 90th, and 100th percentiles, with colors interpolated between anchors.
        (c) Top-10 counties by per-household health cost in 2023.}
       \label{fig:dc_cost_2023_2028}
\end{figure}

We present the total health costs of U.S. data centers in Fig.~\ref{fig:total_health_cost_US_2023_2028}, with details reported Table~\ref{table:history_and_2028_region_health_cost}. In our reference case of scope-1 emissions set at 10\% of permitted levels, scope-2 health costs dominate scope-1 costs. This suggests that, although alternative fuels for on-site diesel generators can help mitigate direct health impacts near data centers, greater health benefits may come from powering data centers with less polluting electricity. However, if scope-1 emissions increase and approach permitted levels
(e.g., due to extended wide-area grid outages), scope-1 health impacts could become substantially larger, especially for data center host communities.

\subsubsection{Uneven Distribution of Data Centers' Public Health Impacts}
The public health impacts of data centers depend largely on the locations of both the data centers and the power plants supplying them. As shown in Fig.~\ref{fig:county_percapita_dc_2023}, health impacts of data centers vary
significantly across counties. The highest county-level per-household health
cost is about seven times the national average and approximately 200 times the
lowest county-level value. This substantial disparity highlights the need to carefully examine local and regional health impacts to support more responsible computing and data center siting.

Importantly, the uneven distribution of per-household health impacts from data
centers is strongly correlated with that of the power grid, with a
correlation coefficient of 0.901. For example, several counties in West
Virginia are among the most affected, reflecting in part the role of
coal-fired power plants in West Virginia in supplying electricity to data
centers in neighboring Virginia. This highlights the need to address data center health disparities
holistically, through both less-polluting power supply and careful data center
siting. 

The county-level total public health impacts of data centers are shown in Appendix~\ref{appendix:county_level_total_2023}. These results suggest that local population density and proximity to emissions sources, including power generation infrastructure, may partly explain why estimated public health impacts are higher in some counties than in others.

\subsubsection{Relative to Fuel Combustion Health Impacts}\label{sec:relative_fuel_combusion}

Stationary-source fuel combustion, including fossil-fuel electricity
generation, industrial and commercial fuel use, and residential wood
combustion, is an important source of anthropogenic air pollutants \cite{Health_COBRA_EPA_Website}. Mobile
sources, such as vehicles, are classified separately in COBRA. Because both scope-1 and scope-2 health impacts of data centers arise from stationary-source fuel combustion, we show the
\emph{relative} health impact of data centers by
considering U.S. fuel combustion as the reference, modeled by COBRA as three broad fuel combustion sectors: electric, industrial, and other \cite{Health_COBRA_EPA_Website}.
Setting emissions from these fuel-combustion sectors to zero in COBRA yields an estimated U.S. fuel-combustion health impact of \$507.79 (\$369.45, \$646.12) billion in 2023.

\begin{wrapfigure}[17]{r}{0.7\textwidth}
\vspace{-0.5cm}
\centering
    \subfloat[Relative health cost]{
    \includegraphics[height=0.22\textwidth,valign=b]{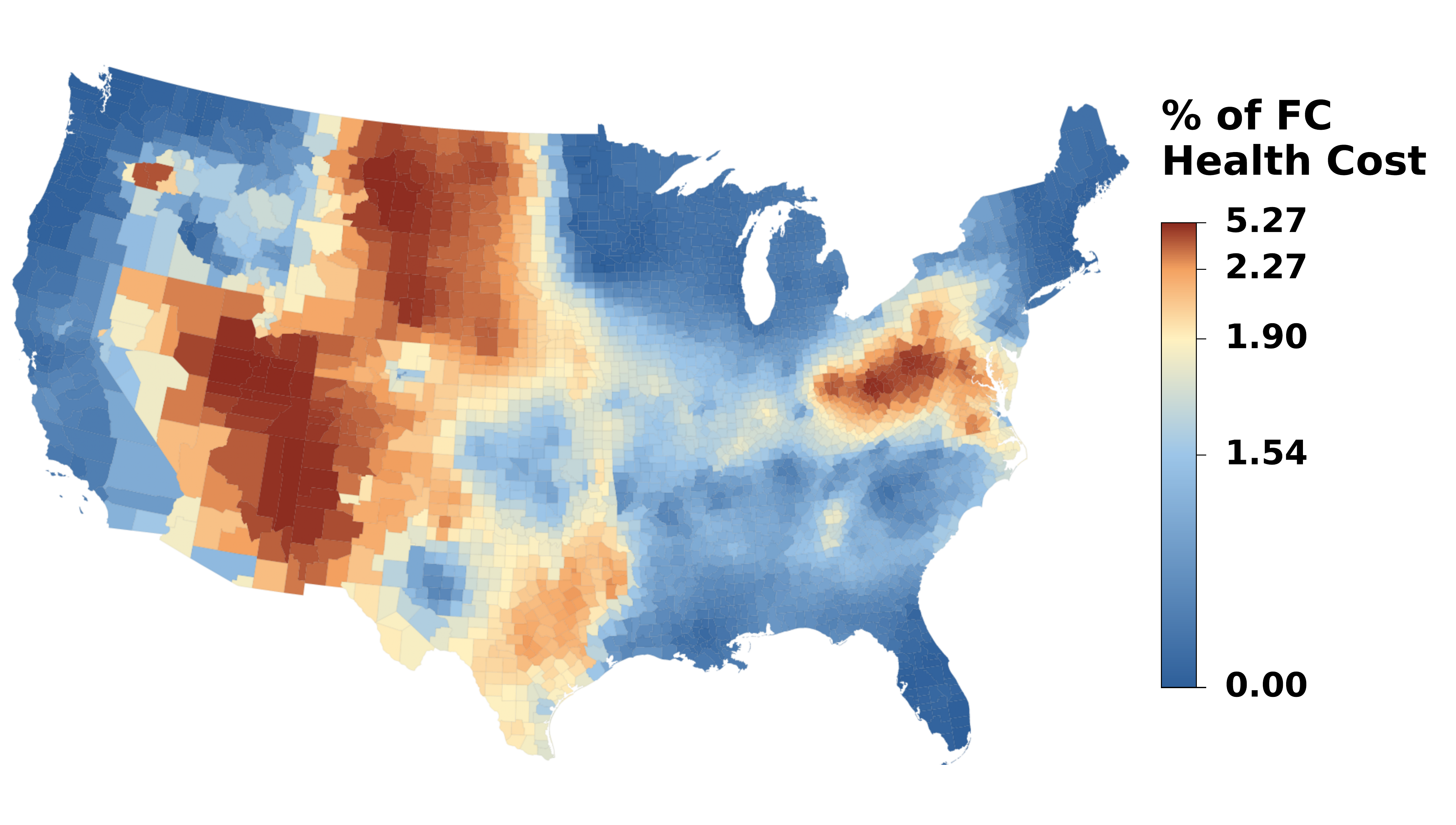} 
    }
    \subfloat[Top-10 counties]{
        \adjustbox{height=0.11\textwidth, valign=b}{
        \centering
        \tiny
        \begin{tabular}{c|c|c} 
        \toprule
        \textbf{State} & \textbf{County} & \begin{tabular}[c]{@{}c@{}}
        \textbf{\% of FC}\\ \textbf{Health Cost}\end{tabular} \\ 
        \hline
        UT & Millard & \textbf{5.27} \\ 
        \hline
        UT & Juab & \textbf{4.91} \\ 
        \hline
        UT & Emery & \textbf{4.52} \\ 
        \hline
        UT & Sanpete & \textbf{4.47} \\ 
        \hline
        UT & Sevier & \textbf{4.36} \\ 
        \hline
        MT & Rosebud & \textbf{4.33} \\ 
        \hline
        UT & Carbon & \textbf{4.24} \\ 
        \hline
        AZ & Apache & \textbf{4.22} \\ 
        \hline
        MT & Custer & \textbf{4.15} \\ 
        \hline
        MT & Garfield & \textbf{3.96} \\ 
        \bottomrule
        \end{tabular}
        }}
        \vspace{-0.1cm}
        \caption{County-level U.S. data center health cost relative to fuel combustion health cost in 2023. The legend shows percentile anchors at the 0th, 50th, 75th, 90th, and 100th percentiles, with colors interpolated between anchors. Per-household data center and fuel combustion health cost correlation coefficient: 0.842. ``FC'' denotes fuel combustion.}\label{fig:county_ratio_dc_all_2023}
\end{wrapfigure}
The ratio of U.S. data center to fuel combustion health costs is approximately 1.31\% in 2023, suggesting a relatively low national burden. 
As shown in Fig.~\ref{fig:county_percapita_dc_2023}, however,
the regional concentration of data center health costs 
underscores the need for closer attention to local impacts.

We analyze the county-level ratio of data center health costs to fuel combustion health costs in Fig.~\ref{fig:county_ratio_dc_all_2023}. The results further indicate a disproportionate distribution of data centers' relative health burden. Moreover, county-level per-household health impacts from data centers are strongly correlated with the fuel combustion health burden, with a correlation coefficient of 0.842. This suggests that data center-related emissions may add to existing public health burdens in some counties already experiencing higher fuel-combustion-related impacts.

While COBRA only provides county-level results, finer-grained analyses, such as at the Census Block level, are an important future direction and can identify opportunities to better target resources toward mitigating the public health impacts of data centers.

\begin{table}[!t]
\centering
\scriptsize
\caption{The public health cost of U.S. data centers in 2023 and projection for 2028. FC: Fuel Combustion}\label{table:history_and_2028_region_health_cost}
\vspace{-0.3cm}
\begin{tabular}{c|c|c|c|c|c|c|c} 
\toprule
\multirow{2}{*}{\textbf{Year}} & 
\multirow{2}{*}{\begin{tabular}[c]{@{}c@{}}\textbf{Electricity} \\(TWh)\end{tabular}} & 
\multirow{2}{*}{\begin{tabular}[c]{@{}c@{}}\textbf{Electricity Cost} \\(billion \$)\end{tabular}} &\multirow{2}{*}{\textbf{Scope}}&\multirow{2}{*}{\textbf{Mortality}} &
\multirow{2}{*}{\begin{tabular}[c]{@{}c@{}}\textbf{Health Cost}\\(billion \$)\end{tabular}} & \multirow{2}{*}{\begin{tabular}[c]{@{}c@{}}\textbf{Per-Household }\\\textbf{Health Cost (\$)}\end{tabular}}& \multirow{2}{*}{\begin{tabular}[c]{c}\textbf{\% of FC}\\\textbf{Health Cost}\end{tabular}}  \\
&&&&&&&\\
\hline

\multirow{3}{*}{\centering 2023}
&\multirow{3}{*}{\centering 176.39}
&\multirow{3}{*}{\centering 15.11}
 &Scope-1&\textbf{32} (26, 37) &\textbf{0.51} (0.43, 0.59) & \textbf{3.65} (3.08, 4.21)&0.10\%\\
\cline{4-8}
 & &&Scope-2&\textbf{401} (294, 508) &\textbf{6.16} (4.59, 7.73) & \textbf{43.83} (32.69, 54.97)&1.21\%\\
\cline{4-8}
 & &&Total&\textbf{433} (320, 546) &\textbf{6.67} (5.03, 8.32) & \textbf{47.48} (35.77, 59.19) &1.31\%\\
\hline

\multirow{3}{*}{\centering 2028 (\textbf{Low})}
&\multirow{3}{*}{\centering 325.00}
&\multirow{3}{*}{\centering 27.84}
 &Scope-1&\textbf{54} (46, 63)&\textbf{0.90} (0.77, 1.03)&\textbf{6.11} (5.22, 7.00) &0.15\%\\
\cline{4-8}
 & &&Scope-2&\textbf{650} (483, 818) &\textbf{10.78} (8.14, 13.41) &\textbf{73.29} (55.37, 91.21)&1.82\%\\
\cline{4-8}
 & &&Total&\textbf{705} (529, 880)&\textbf{11.67} (8.91, 14.44) & \textbf{79.40} (60.59, 98.21)&1.97\%\\
\hline

\multirow{3}{*}{\centering 2028 (\textbf{High})}
&\multirow{3}{*}{\centering 580.00}
&\multirow{3}{*}{\centering 49.68}
 &Scope-1&\textbf{97} (82, 112) &\textbf{1.61} (1.37, 1.84) &\textbf{10.94} (9.35, 12.53) &0.27\%\\
\cline{4-8}
 & &&Scope-2&\textbf{1165} (865, 1464)&\textbf{19.29} (14.58, 24.01) &\textbf{131.23} (99.14, 163.31) &3.26\%\\
\cline{4-8}
 & &&Total&\textbf{1262} (947, 1576) &\textbf{20.90} (15.95, 25.85) &\textbf{142.16} (108.49, 175.84)&3.53\%\\
\bottomrule

\end{tabular}
\end{table}

\subsection{Projection for 2028}

As shown in Table~\ref{table:history_and_2028_region_health_cost}, the growing demand for data centers is projected to increase their public health impacts, reaching \$11.7 billion and \$20.9 billion in 2028 under the low- and high-growth scenarios, respectively. These estimates represent 75\% and 213\% increases relative to the 2023
level, respectively, while U.S. fuel-combustion-related health costs are
projected to increase by 17\% over the same period.

Under the high-growth scenario, the estimated health outcomes associated with U.S. data center emissions in 2028 include approximately 600,000 asthma symptom cases and 1,300 premature deaths, along with other morbidity outcomes. 
While this impact remains relatively small compared with the overall national public health burden and the broad stationary fuel-combustion sectors,  the projected growth, localized concentration, and uneven geographic distribution of the health costs warrant closer attention.

\section{Health-Informed Computing: Addressing Data Centers' Public Health Impact}

In this section, we present Health-Informed Computing, a framework that explicitly incorporates public health impacts as a key optimization objective and strategically manages data center workloads to reduce adverse health outcomes while supporting broader sustainability goals.

\subsection{Spatiotemporal Heterogeneity of Health Prices}

The grid fuel mix varies substantially across space and time \cite{WattTime_Website,Google_CarbonAwareComputing_PowerSystems_2023_9770383}, depending on generation schedules, transmission constraints, electricity demand, and other factors. Because power plants and fuels differ in their air pollutant emission rates, variation in the grid fuel mix creates spatiotemporal heterogeneity in the externalized scope-2 public health impact of electricity use, which we refer to as the \emph{health price}, measured in dollars per unit of electricity consumed. 

Although criteria air pollutants can share common sources with carbon emissions, health price may not be well captured by carbon intensity alone.
Carbon emissions have approximately the same climate impact regardless of where they are emitted. By contrast, the same amount of criteria air pollutants can lead to very different health impacts depending on the emitting source, atmospheric conditions, downwind transport, baseline air quality, and the size of exposed populations.
 
As a result,  low-carbon electricity does not necessarily imply a low public health burden, and vice versa. For example, two regions with similar carbon intensities may have different health costs if their generators are located near populations with different exposure risks or if local weather conditions affect pollutant dispersion differently. Conversely, regions with different carbon intensities may produce similar health impacts if their air pollutant emissions, atmospheric conditions and exposed populations are comparable.

\begin{figure}[!t]
\centering
\subfloat[]{
 \includegraphics[width=0.32\textwidth,valign=b]{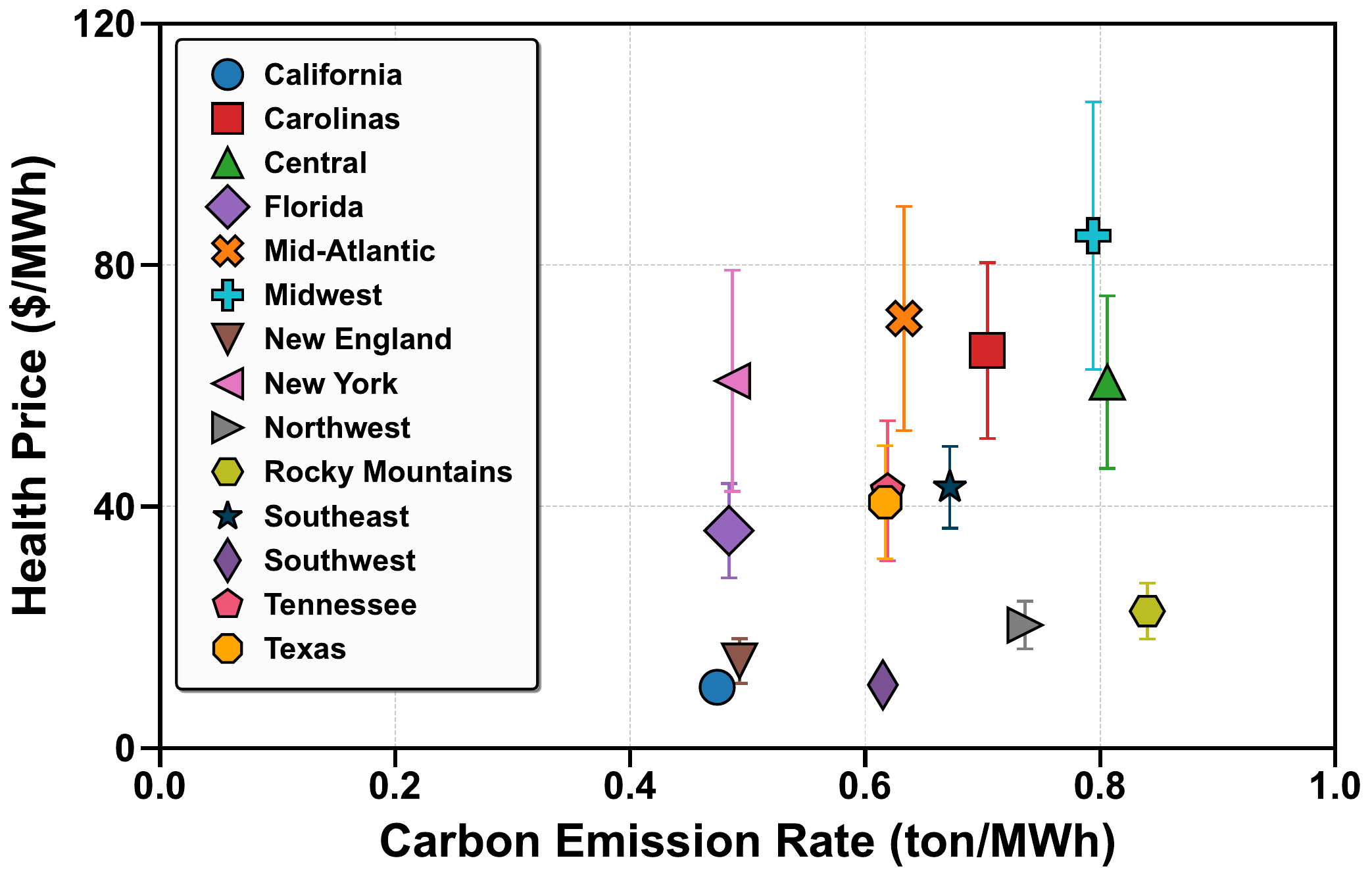}
 \label{fig:health_carbon_corr_epa_2023}
}
\subfloat[]{
\includegraphics[width=0.32\textwidth,valign=b]{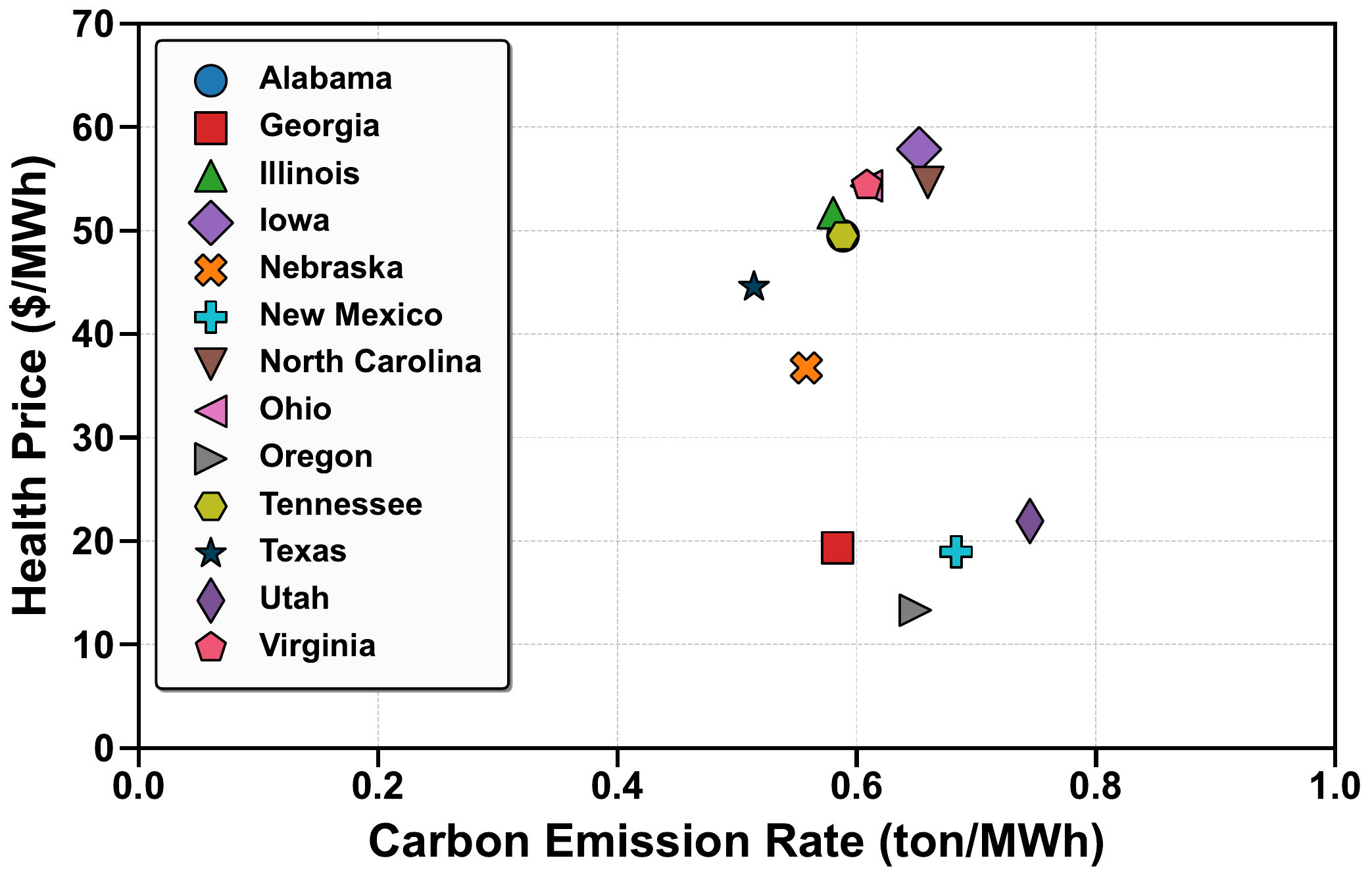}\label{fig:glb_corr_plot}
}
\subfloat[]{
\includegraphics[width=0.32\textwidth,valign=b]{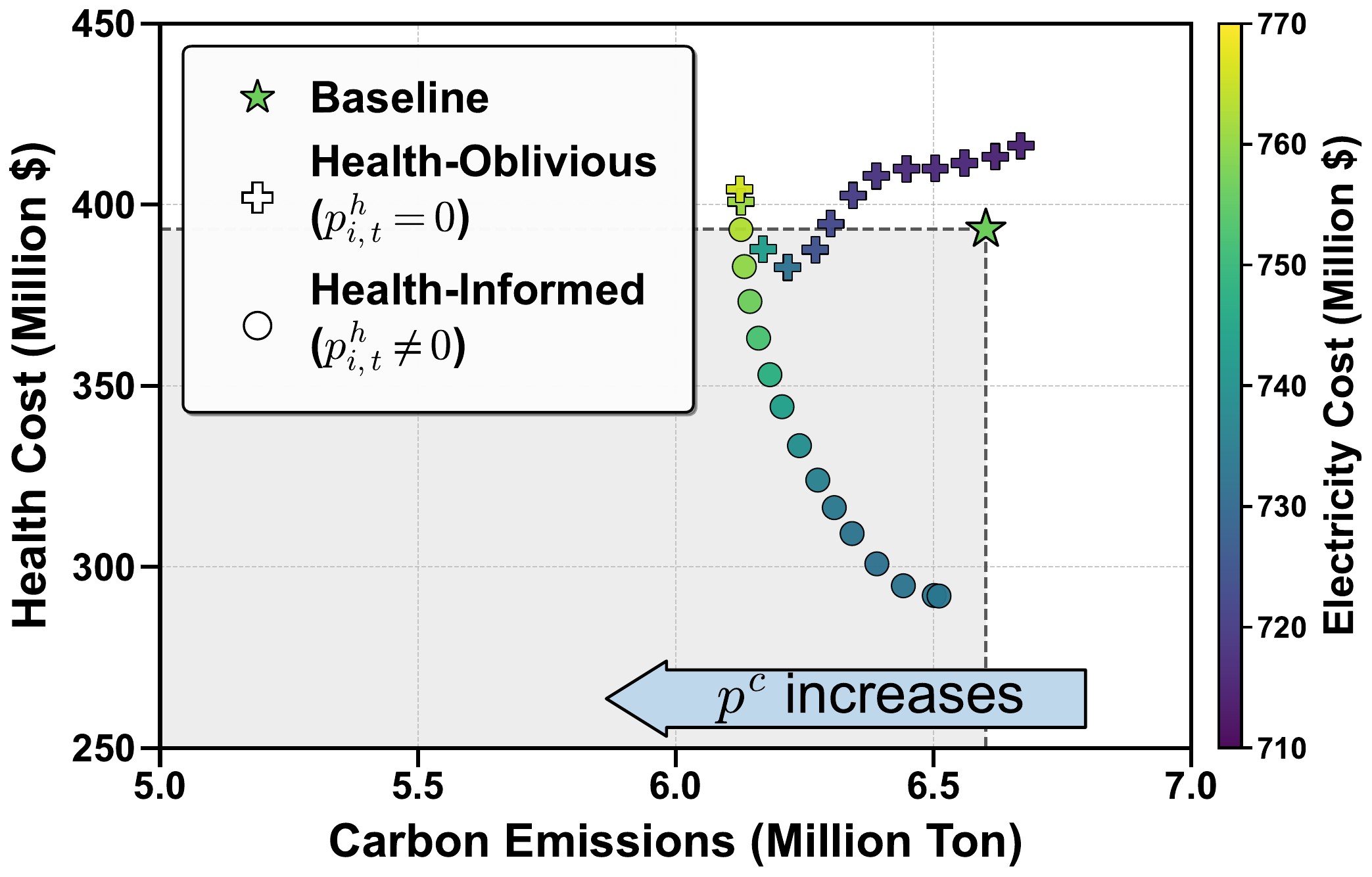}\label{fig:glb_tradeoff_plot}
}
\vspace{-0.3cm}
\caption{Marginal health impact vs. marginal carbon emission
based on the 2023 data.
(a) U.S. EPA signals for 14 U.S. independent electricity regions \cite{Health_AVERT_Emission_Marginal_EPA_Website,Health_Benefit_kWh_EPA_Website}. The error bar indicates
the high and low health prices. Carbon and (mid) health
correlation coefficient:  0.338.
(b) WattTime signals for
a large technology company's U.S. data center locations
\cite{WattTime_Website,Facebook_SustainabilityReport_2024}.
 Carbon and health
correlation coefficient:  -0.351.
(c) Results for different GLB settings ($p^c$ varying from 0 to $\infty$) using a large technology company's 2023 energy usage across its U.S. data center locations, assuming that each data center can consume up to
1.5 times its baseline energy use.}\label{fig:glb_case_study}
\end{figure}

To further illustrate this point, we analyze the marginal health price and marginal carbon emission rate from the U.S. EPA's AVERT model across the 14 grid regions in 2023 \cite{Health_AVERT_Emission_Marginal_EPA_Website,Health_Benefit_kWh_EPA_Website}. We use the U.S. EPA's ``uniform EE'' scenario for health price
\cite{Health_Benefit_kWh_EPA_Website}, which estimates the marginal health impact of a constant load. As shown in Fig.~\ref{fig:health_carbon_corr_epa_2023}, health price exhibits greater spatial and temporal variability than carbon intensity, with a weak-to-moderate spatial correlation between the two signals (correlation coefficient 0.338). 
We also analyze 5-minute marginal health price and carbon intensity signals across the U.S. from October 1, 2023, to September 30, 2024, using data from WattTime~\cite{WattTime_Website}.
The details are provided in Appendix~\ref{appendix:health_informed_opportunity}. The results show similar patterns, reaffirming that health prices can differ significantly from carbon intensities.

Our analysis suggests that carbon-aware computing alone may miss important opportunities to mitigate public health impacts, motivating the development of health-informed computing.

\subsection{Health-Informed Computing}

Computing workloads in data centers often have substantial scheduling flexibility \cite{Google_CarbonAwareComputing_PowerSystems_2023_9770383}.  For example,
as supported by EPRI's recent initiative
on maximizing data center load flexibility \cite{DataCenter_FlexibilityInitiative_EPRI_WhitePaper_2024}, 
AI training can be shifted across data centers or paused temporarily. 
When combined with the spatiotemporal heterogeneity of scope-2 health prices, this flexibility creates opportunities to reduce public health impacts by shifting computation toward data centers and/or hours with lower health costs.

To date, data center scheduling flexibility has been used primarily to reduce electricity costs, carbon emissions~\cite{Google_CarbonAwareComputing_PowerSystems_2023_9770383}, water consumption~\cite{Water_StressWeighted_SustainableComputing_DingYi_2025_10.1145/3757892.3757904},  environmental inequity~\cite{Shaolei_Equity_GLB_Environmental_AI_eEnergy_2024},
and emergency demand response~\cite{Shaolei_Colocation_EDR_Adam_Performance_2015}. We propose \emph{Health-Informed Computing} (\ouralg), which builds on existing scheduling systems and incorporates public health impact as an additional metric to prioritize
data centers with lower health impacts.

We use geographical load balancing (GLB) as a concrete example.
A canonical objective in GLB is carbon-aware computing, which dynamically
dispatches workloads to data centers with lower carbon intensities
\cite{Google_Demand_Response_Carbon_2025}.
To further incorporate public health impacts, the GLB objective for each time slot
$t$ can be written as:
\begin{equation}\label{eqn:health_GLB}
    \min_{\mathbf{w}_{t}\in\mathcal{W}_t}\sum_{i = 1}^N \left(p_{i,t}^e + p_{i,t}^h + p^c r_{i,t}^c\right)\cdot w_{i,t},
\end{equation}
where $N$ is the number of data centers,
$p_{i,t}^e$, $p_{i,t}^h$, and $r_{i,t}^c$ denote, respectively,
the electricity price, health cost, and carbon emissions per kWh at
data center $i$ and time $t$,
$p^c$ denotes the carbon price,
$\mathbf{w}_{t}=(w_{1,t},\cdots,w_{N,t})$ denotes the energy use induced by
the dispatched workloads, and
$\mathcal{W}_t$ denotes the feasible set at time $t$, capturing operational
constraints such as latency constraints and workload-specific requirements \cite{Gao:2012:EG:2377677.2377719}.
 Carbon prices commonly fall in the range of \$10 to \$200 per ton, depending on the market and compliance requirements
\cite{DataCenter_Health_25Billion_CarbonIncluded_CMU_Paper_2026_Muller2026DataCentersPollutionGHG}.

By adding the health price $p_{i,t}^h$, the GLB algorithm optimizes a more
holistic objective that accounts for electricity costs, carbon emissions, and
public health impacts. All else being equal, this objective prioritizes data
centers with lower health impacts. Importantly, the objective in~\eqref{eqn:health_GLB} recovers existing carbon-aware GLB when the health price $p_{i,t}^h = 0$. 

At each time slot, the scheduler considers electricity price, carbon cost, and public health price across data centers, subject to practical constraints such as capacity, latency, workload-specific requirements, and data availability. Flexible workloads can then be shifted toward locations and times with a lower combined cost as formulated in \eqref{eqn:health_GLB}.

Beyond GLB, \ouralg can help data centers reduce externalized public health costs and better align operational decisions with corporate responsibility and community-impact goals. It also provides a practical mechanism for incorporating public health considerations into data center siting and energy procurement decisions. At the same time, broader deployment of \ouralg requires more accurate, timely, and standardized public health impact data, which remains an important direction for future research.

\begin{table*}[!t]
\centering
\caption{\ouralg under different settings ($\lambda=1.5$), with percent changes
in parentheses reported relative to the baseline.
Three special cases:
Electricity Cost Optimal (\textbf{\eopt}, with $p^c=0$ and $p^h_{i,t}=0$),
Carbon Emission Optimal (\textbf{\copt}, with $p^e_{i,t}=0$ and $p^h_{i,t}=0$),
and Health Impact Optimal (\textbf{\hopt}, with $p^e_{i,t}=0$ and $p^c=0$).
} 
\vspace{-0.3cm}
\label{table_results_glb}
\scriptsize
\setlength{\tabcolsep}{3pt}
\renewcommand{\arraystretch}{1.15}
\resizebox{\textwidth}{!}{%
\begin{tabular}{c|c|cc|cc|cc|cc|cc|cc|cc|cc}
\toprule
\multirow{2}{*}{\textbf{Metric}} &
\multirow{2}{*}{Baseline} &
\multicolumn{2}{c|}{\multirow{2}{*}{\eopt}} &
\multicolumn{2}{c|}{\multirow{2}{*}{\copt}} &
\multicolumn{2}{c|}{\multirow{2}{*}{\hopt}} &
\multicolumn{4}{c|}{Health-Oblivious ($p^h_{i,t}=0$)} &
\multicolumn{6}{c}{Health-Informed ($p^h_{i,t}\not=0$)} \\
\cline{9-18}
&& \multicolumn{2}{c|}{} &
\multicolumn{2}{c|}{} &
\multicolumn{2}{c|}{} &
\multicolumn{2}{c|}{$p^c=$\$50/ton} &
\multicolumn{2}{c|}{$p^c=$\$200/ton} &
\multicolumn{2}{c|}{$p^c=$\$0/ton} &
\multicolumn{2}{c|}{$p^c=$\$50/ton} &
\multicolumn{2}{c}{$p^c=$\$200/ton} \\
\hline
Health (Million \$) &
393.23 &
416.29 & (+5.86\%) &
404.26 & (+2.80\%) &
\textbf{289.67} & \textbf{(-26.34\%)} &
400.81 & (+1.93\%) &
383.32 & (-2.52\%) &
291.94 & (-25.76\%) &
294.21 & (-25.18\%) &
305.99 & (-22.18\%) \\
\hline
Energy (Million \$) &
756.50 &
\textbf{714.99} & \textbf{(-5.49\%)} &
765.68 & (+1.21\%) &
741.49 & (-1.98\%) &
720.49 & (-4.76\%) &
734.77 & (-2.87\%) &
733.66 & (-3.02\%) &
732.69 & (-3.15\%) &
732.11 & (-3.22\%) \\
\hline
Carbon (Million Ton) &
6.60 &
6.67 & (+1.02\%) &
\textbf{6.12} & \textbf{(-7.23\%)} &
6.54 & (-0.89\%) &
6.33 & (-4.07\%) &
6.20 & (-6.01\%) &
6.51 & (-1.38\%) &
6.45 & (-2.31\%) &
6.36 & (-3.67\%) \\
\bottomrule
\end{tabular}
}
\end{table*}

\subsection{Empirical Evaluation}

To illustrate the empirical impact of \ouralg, we evaluate
GLB under a standard formulation with workload-conservation and per-site
capacity constraints. The evaluation is based on a technology company's U.S.
data center locations and uses its published 2023 per-site energy consumption
as the baseline workload assignment
\cite{Facebook_SustainabilityReport_2024}.
Because the U.S. EPA provides health prices only on an annualized basis,
we obtain 5-minute health prices and carbon emission rates from WattTime
\cite{WattTime_Website}. Note that WattTime also uses a simplified S-R matrix for air dispersion modeling, but unlike COBRA which models a variety of health outcomes, it considers only mortality in its health price estimate. 
Importantly, because health prices involve uncertainties from air-dispersion
modeling and health-outcome estimation, our results should be interpreted as
illustrative rather than as an actual health impact assessment of the
technology company's public health impacts.

We vary the carbon price parameter $p^c\in[0,\infty]$ to characterize how different
scheduling objectives trade off electricity cost, carbon emissions, and public
health impacts. The resulting curve is shown in
Fig.~\ref{fig:glb_tradeoff_plot}, with detailed results for representative settings
reported in Table~\ref{table_results_glb}.

Specifically, Fig.~\ref{fig:glb_tradeoff_plot} shows that, compared with the
health-oblivious case (i.e., $p^h_{i,t}=0$), \ouralg can substantially reduce
public health impacts while keeping carbon emissions and electricity costs
within a moderate range. This highlights that explicitly accounting for health
impacts in GLB can uncover scheduling opportunities that are not captured by
existing carbon-aware computing alone.
Compared with the baseline, \ouralg reduces the estimated health cost from about
\$390 million to below \$300 million, while maintaining a similar level of
carbon emissions. Notably, as the carbon price $p^c$ increases, \ouralg traces a clear
tradeoff curve: prioritizing carbon reductions lowers emissions but may 
increase health and electricity costs. 

The results can be further explained by the relationship between health prices
and carbon emission rates across the technology company's 13 U.S. data center
locations used in our evaluation, as shown in Fig.~\ref{fig:glb_corr_plot}.
The correlation coefficient is approximately $-0.351$, indicating a
weak-to-modest negative association between the two metrics in this setting. Thus,
locations with lower carbon emissions do not necessarily have lower public
health impacts. Moreover, consistent with the U.S. EPA's nationwide analysis in
Fig.~\ref{fig:health_carbon_corr_epa_2023}, health prices exhibit greater
spatial variability than carbon emission rates across the data center
locations. This helps explain why \ouralg can substantially reduce public
health impacts while only modestly affecting carbon emissions.

\section{Additional Recommendations}\label{sec:recommendation}

We provide the following additional recommendations
to complement \ouralg.

\textbf{Recommendation 1: Extended Energy Reporting.}
Despite their immediate public health impacts, criteria air pollutants and their downstream health effects are rarely included in technology companies' sustainability reports. 
Importantly, these impacts do not necessarily require a separate reporting framework: much of the needed information, including data center location, electricity use, and on-site fuel use, already overlaps with energy reporting.
We therefore recommend extending existing energy reporting practices to include criteria air pollutants and associated public health impacts across regions.

\textbf{Recommendation 2: Attention to All.}
While data centers' scope-1 health impacts on host communities are increasingly
recognized, scope-2 impacts on other affected communities have received 
less attention. This gap may make it harder to identify and address communities affected by scope-2 emissions.
To support more comprehensive sustainability assessment,
we recommend that technology companies evaluate the cross-state public health burdens of
their operations when deciding where to build data centers and how to source
electricity.

\textbf{Recommendation 3: Improving Data Accuracy for Health Impact Assessment.}
To estimate data centers' public health impacts with greater accuracy and better inform potential mitigation strategies, we recommend further interdisciplinary research that integrates air-quality dispersion modeling, health economics, epidemiology, and health-informed computing. Such research can better capture cross-state and downwind impacts, improve estimates of population-level morbidity and mortality, and translate these estimates into actionable strategies for siting, energy procurement, workload scheduling, and backup-generation management.

\section{Conclusion}\label{sec:conclusion}

In this paper, we 
model U.S. data centers' air pollutant emissions and estimate their associated public health impacts using COBRA.
Our findings suggest that the total annual public health burden of U.S. data centers could exceed \$20 billion in 2028 under the high-growth scenario. Importantly, these health costs are not evenly distributed, with certain communities bearing a disproportionate share. 

We also propose \ouralg, a novel framework that explicitly incorporates public health impact as a key metric when siting data centers and/or scheduling computing workloads. 
More broadly, we recommend extended energy reporting to account for the public health impacts of data centers and paying attention to all impacted communities, thereby supporting responsible and sustainable development of future computing infrastructures.

\section*{Acknowledgement}
The authors would like to thank Susan Wierman for providing comments on an initial draft of this paper.

{
       \bibliographystyle{unsrt}

}

\newpage
\appendix 
\section*{Appendix}

\section{Modeling Details}\label{sec:methodology_details}

We describe the evaluation methodology used for our empirical analysis. 
We use the COBRA Desktop v5.1 provided by the U.S. EPA \cite{Health_COBRA_EPA_Website} to study the public health impact of U.S. data centers in both 2023 and 2028. While COBRA uses a reduced-complexity air quality dispersion model based on a source-receptor matrix for rapid evaluation, its accuracy has been validated and the same or similar model has been commonly adopted in the literature for large-area air quality and health impact analysis \cite{EPA_Papers_Cite_COBRA_Website,Health_BitCoin_HarvardPublicHealth_WattTime_preprint_2024_guidi2024environmental,Health_InMAP_AirQuality_Model_2017_tessum2017inmap,Health_DeathImport_Energy_US_700_California_Stanford_ENvironmentalResearch_2022_hennessy2022distributional}.
We consider county-level air pollutant dispersion throughout the contiguous U.S., which is the area currently supported by COBRA \cite{Health_COBRA_EPA_Website}. Note that cities considered
county-equivalents for census purposes are also referred to as ``counties''
in COBRA. Throughout the paper, we use ``county'' without further specification. 

All the monetary values are presented in the 2023 U.S. dollars unless otherwise stated. We set
the discount rate as 2\% in COBRA as recommended by the EPA
based on the U.S. Office of Management and Budget
Circular No. A-4 guidance \cite{Health_COBRA_EPA_Website}.
When presenting a single value or a ratio (e.g., health-to-electricity cost ratio) if applicable, we use the midrange of the low and high estimates provided by COBRA.

 COBRA provides data for county-level population, health
incidence, valuation, and baseline emissions 
for 2023 and 2028
\cite{Health_COBRA_EPA_Website}. 
Specifically, the baseline emissions used by COBRA are based on the U.S. EPA's Emissions Modeling Platform 2016v1 and account for federal and state regulations as of May 2018 \cite{Health_COBRA_EPA_Website}. Regulatory and emissions changes since then may affect the accuracy of the estimates. Therefore, the modeled health impacts should be interpreted as a first-order approximation, and future re-analysis using updated dispersion models, emissions data, population data, and health economics assumptions would help improve the assessment.

\textbf{Electricity price.}
When estimating the electricity cost for data centers in 2023 and 2028, we use the state-level average price for industrial users in \cite{EIA_ElectricPower_Annual_2023}. 
The projected U.S. nominal electricity price for industrial users remains nearly the same from 2023 to 2030 (24.96 \$/MMBtu in 2023 vs. 23.04 \$/MMBTu in 2030) in the baseline case per the EIA's Energy Outlook 2023 \cite{EIA_EnergyOutlook_2023_website}. Thus, our estimated health-to-electricity cost ratio may be even higher if we further adjust inflation.

\subsection{Public Health Impact of On-road Vehicles}\label{appendix:on_road}

\begin{wrapfigure}[18]{r}{.4\textwidth}
	\centering
 \vspace{-0.1cm}	
    \centering
\includegraphics[width=1\linewidth]{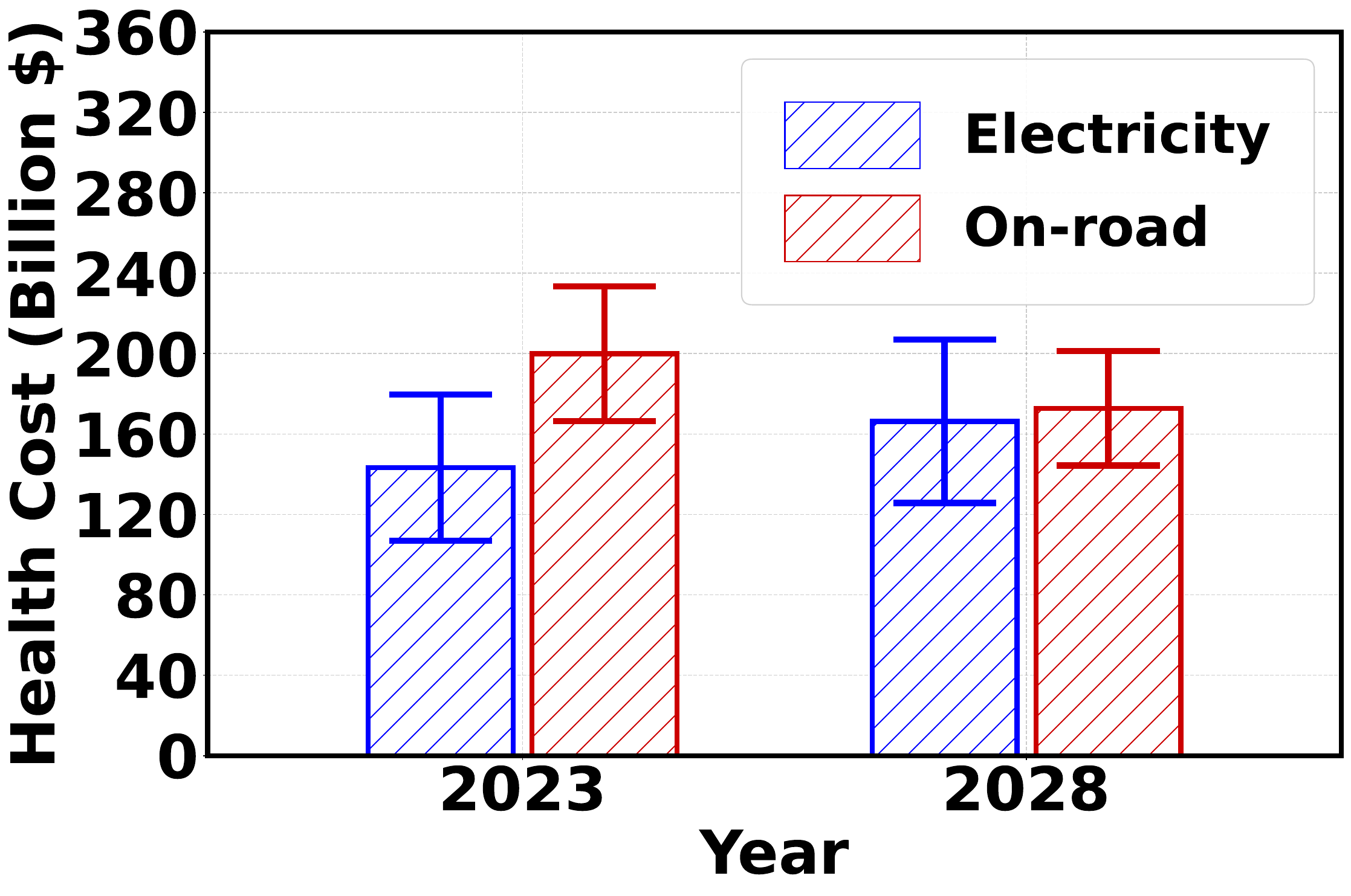}
 \vspace{-0.8cm}	
    \caption{Public health costs of electricity generation and on-road emissions in the contiguous U.S. in 2023 and 2028 \cite{Health_COBRA_EPA_Website}. The error bars represent high and low estimates returned by COBRA using two different exposure-response functions.}
    \label{fig:on_road_comparsion}
\end{wrapfigure}
Mobile sources, including vehicles, marine engines, and generators, collectively
are a major contributor to air pollution in the U.S., with automobile (vehicles) being a primary contributor \cite{Health_Vehicle_Emission_PrimarySource_US_NationalParkService_Website,EPA_Mobile_Source_MoreThanHalf_Pollutants_Website}. 
 In the COBRA model, on-road emissions are categorized under the ``Highway Vehicles'' sector and include both tailpipe exhaust and tire and brake wear. 
Based on the emission data projected by 
the U.S. EPA's COBRA modeling tool \cite{Health_COBRA_EPA_Website},
we show in  Fig.~\ref{fig:on_road_comparsion} that the electric power sector's total public health cost  in the contiguous U.S. approaches the scale of on-road vehicle emissions by all the registered vehicles (including tailpipe exhausts and brakes) in 2028.

Addressing air-pollution-related health impacts requires coordinated efforts across sectors, and emissions from the electric power sector and on-road vehicles call for different mitigation strategies. Therefore, our comparison between the electric power sector impacts and on-road emissions is intended only to provide a sense of scale. It should not be interpreted as suggesting that one sector is more or less important than another, or that mitigation approaches are directly interchangeable across sectors.

\subsection{Public Health Impact of Backup Generators in Virginia}\label{sec:backup_generator_northern_virginia}

We collect a dataset of the air quality permits:
permits issued before January 1, 2023, from \cite{Air_DataCenter_Diesel_Generator_PiedmontEnvironmentalCouncil_Web_Map}, and permits issued between January 1, 2023 and December 1, 2024,
from \cite{Virginia_AirPermitsDataCenter_Website}. 
The total permitted site-level annual emission limits   
are approximately
13,000 tons of \nox, 1,400 tons of VOCs, 50 tons of \sotwo,
and 600 tons of \pmtwo, all in U.S. short tons.
Assuming actual emissions equal 10\% of the permitted level as of December 2024 as a reference case, our estimates indicate that backup generators could contribute to approximately 14,000 asthma symptom cases and 13--19 premature deaths each year, among other health impacts. These impacts correspond to an estimated annual public health burden of \$220--300 million across the U.S., including \$190--260 million in Virginia, West Virginia, Maryland, Pennsylvania, New York, New Jersey, Delaware, and Washington, D.C., based on COBRA modeling under the ``Fuel Combustion: Industrial'' sector.
If the data center diesel generators in Northern Virginia emit air pollutants at the maximum permitted levels during a prolonged regional grid outage, the emission of \nox could exceed
half of the annual total emissions by all sources in the region \cite{Virginia_AirPermitsDataCenter_Report_JLARC_2024}.
In this hypothetical upper-bound scenario, the estimated annual public health cost would be \$2.2-3.0 billion.

\begin{figure}[h!]
    \centering
   \subfloat[County-level health cost]{\includegraphics[height=0.235\textwidth,valign=b,trim={18cm 6cm 0 0cm},clip]{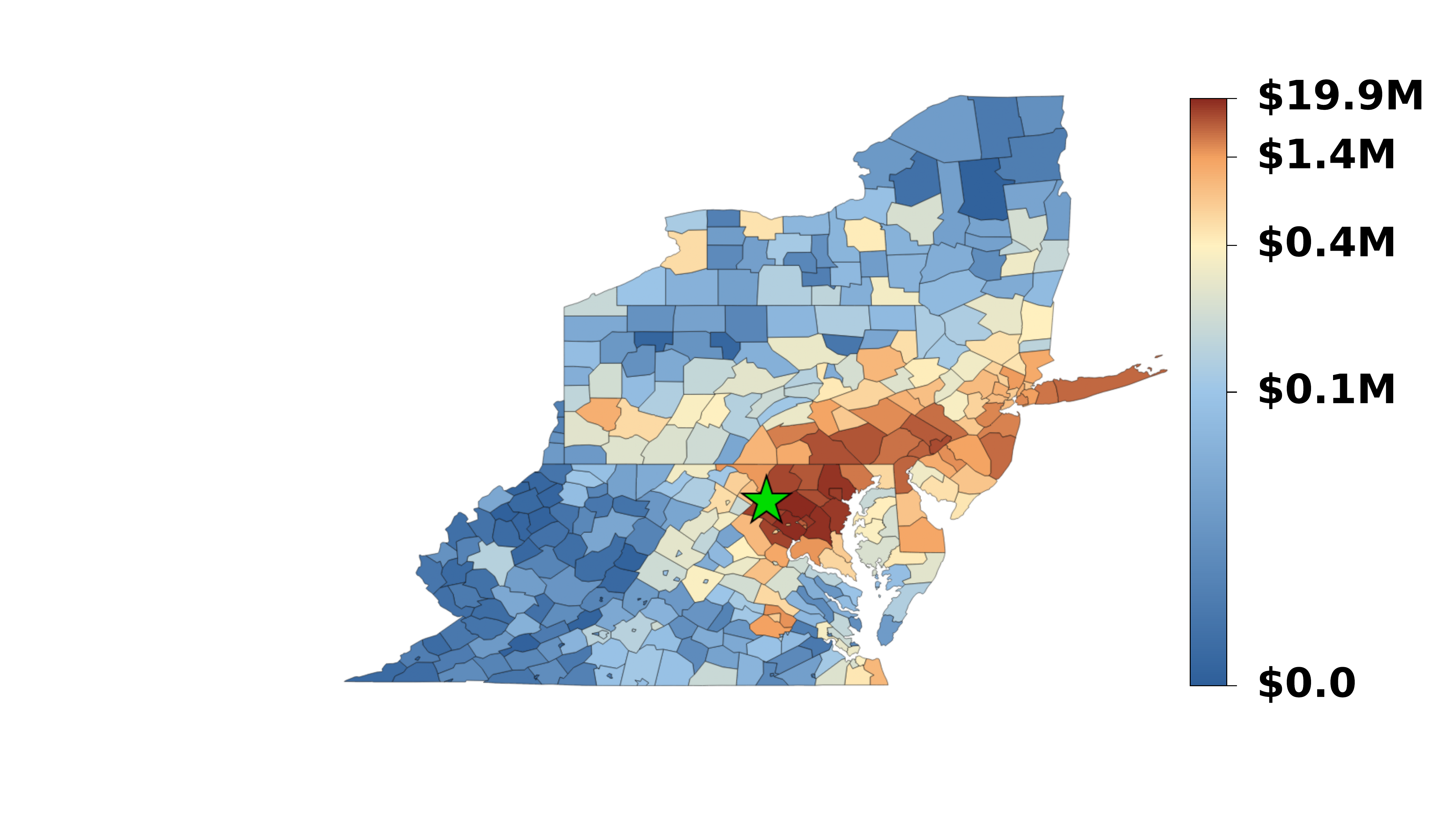}\label{fig:generator_VA_health_2023/map_generator_health}}
    \subfloat[Top-10 counties by health cost]{
        \adjustbox{height=0.11\textwidth,valign=b}{
        \centering
        \scriptsize
        \begin{tabular}{c|c|c} 
        \toprule
        \textbf{State} & \textbf{County} & \begin{tabular}[c]{@{}c@{}}\textbf{Health Cost }\\(million \$)\end{tabular} \\ 
        \hline
        MD & Montgomery & \textbf{19.9} (17.3, 22.4) \\ 
        \hline
        VA & Fairfax & \textbf{18.9} (16.6, 21.2) \\ 
        \hline
        MD & Prince Georges & \textbf{8.9} (7.5, 10.4) \\ 
        \hline
        MD & Baltimore & \textbf{8.3} (7.0, 9.6) \\ 
        \hline
        DC & District of Columbia & \textbf{7.6} (6.2, 9.0) \\ 
        \hline
        MD & Anne Arundel & \textbf{6.3} (5.5, 7.2) \\ 
        \hline
        MD & Baltimore City & \textbf{6.0} (4.8, 7.1) \\ 
        \hline
        VA & Loudoun & \textbf{5.4} (4.7, 6.1) \\
        \hline
        VA & Prince William & \textbf{5.0} (4.4, 5.7) \\ 
        \hline
        MD & Frederick & \textbf{4.6} (3.9, 5.2) \\ 
        \bottomrule
        \end{tabular}
        }}\\ 
        \vspace{-0.3cm}
        \caption{County-level scope-1 health costs from data center backup generators in Virginia under the reference case, assuming emissions equal 10\% of total permitted emissions as of December 2024.    
(a) 
Choropleth map. Loudoun County, Virginia, which has the largest concentration of data centers, is marked in a green star. The legend shows percentile anchors at the 0th, 50th, 75th, 90th, and 100th percentiles, with colors interpolated between anchors.
 (b) Top-10 counties by the total health cost.}
\label{fig:backup_generator_northern_VA_subset}
\end{figure}

We show the county-level health cost and the top 10 counties in Figure~\ref{fig:backup_generator_northern_VA_subset}. The results suggest that elevated health impacts can occur not only in host communities, but also in downwind communities, including those far from the data centers, due to long-range pollutant transport. In addition, higher population density can amplify aggregate health costs, even when per-household impacts are relatively lower.

\subsection{Public Health Impact of a Semiconductor Facility}\label{sec:semiconductor_calculation}

Although semiconductor manufacturing facilities are subject to air quality regulations \cite{EPA_Semiconductor_Rules_Website}, their emissions can still contribute to regional air quality and public-health concerns, depending on location, emissions profile, and exposed populations. 

We consider a semiconductor manufacturing facility located
in Ocotillo, a neighborhood in Chandler, Arizona \cite{Intel_Arizona_OcotilloFacility_Website}. By averaging
the rolling 12-month air pollutant emission levels listed
in the recent air quality monitoring report
(as of October, 2024) \cite{Intel_Arizona_OcotilloFacility_Report_2024}, 
we obtain the annual emissions as follows:
150.4 tons of \nox, 82.7 tons of VOCs,
1.1 tons of \sotwo, and 28.9 tons of \pmtwo.
By applying these on-site emissions to COBRA under the
``Other Industrial Processes'' sector, 
we obtain a total public health cost
of \$14-21 million. Additionally, the total annual energy consumption 
by the facility
is 2074.88 million kWh as of Q2, 2024 \cite{Intel_Arizona_OcotilloFacility_Website}.
Assuming 84.2\% of the energy comes from the electricity
based on the company's global average \cite{Intel_Sustainability_Report_2023_24},
we obtain the facility's annual electricity consumption 
as 1746.63 million kWh. By using the average attribution method, we
further obtain an estimated health cost 
of \$12-17 million
associated with the electricity consumption.
Thus, the total health cost of the facility is \$26-39 million.
This example is illustrative and is not included in the national data-center public-health estimates reported in our main analysis.

By relocating the facility from Chandler, Arizona, to
a planned site in Licking County, Ohio, and assuming the same emission level
and electricity consumption in a hypothetical scenario, 
we can obtain
the total health cost of \$94-156 million, including 
 \$23-36 million attributed to direct on-site emissions and \$70-120 million
attributed to electricity consumption.

\subsection{Energy Consumption for Training a Generative AI Model}

We consider Llama-3.1 as an example generative AI model.
According to the model card \cite{Llama_31_ModelCard_GitHub_2024},
the training process of Llama-3.1 (including 8B, 70B, and 405B)
utilizes 
 a cumulative of 39.3 million GPU hours of computation on H100-80GB hardware, and each GPU has a thermal design power of 700 watts.
Considering
Meta's 2023 PUE of 1.08 \cite{Facebook_SustainabilityReport_2024}
and excluding the non-GPU overhead
for servers, we estimate the total training energy consumption as approximately
30 GWh. Our estimation method follows Meta’s guideline \cite{Llama_31_ModelCard_GitHub_2024}.
It excludes non-GPU server overheads such as CPUs and memory, but actual training electricity use may differ depending on utilization, power management, infrastructure overhead, and accounting boundaries.

\subsection{Average Emission for Each LA-NYC Round Trip 
 by Car}\label{sec:car_emission_la_nyc}

We use the 2023 national average emission rate for light-duty vehicles (gasoline) provided
by the U.S. Department of Transportation  \cite{Health_VehicleEmissionRates_US_DoT_TransportationStatistics_June_2024}.
  The emission rate accounts for tailpipe exhaust, tire wear and brake wear. 
Specifically, 
the average \pmtwo emission rate is
0.008 grams/mile (including
0.004 grams/mile for exhaust,
0.003 grams/mile for brake wear,
and 0.001 grams/mile for tire wear),
and the average \nox emission rate is
0.199 grams/mile for exhaust.
We see that half of \pmtwo for light-duty vehicles comes from
brake and tire wear (0.004 gram/miles), which are also produced
by other types of vehicles including electric vehicles.
The distance for a round-trip between Los Angeles, California,
and New York City, New York, is about 5,580 miles.
Thus, the average auto emissions for each LA-NYC round trip
are estimated as 44.64 grams of \pmtwo and 1110.42 grams of \nox.

\subsection{State-wide Electricity Consumption by U.S. Data Centers in 2023}\label{sec:additional_results_2023}

\begin{figure}[!ht]
    \centering
    \subfloat[Electricity consumption map]{
    \includegraphics[height=0.23\textwidth,valign=b]{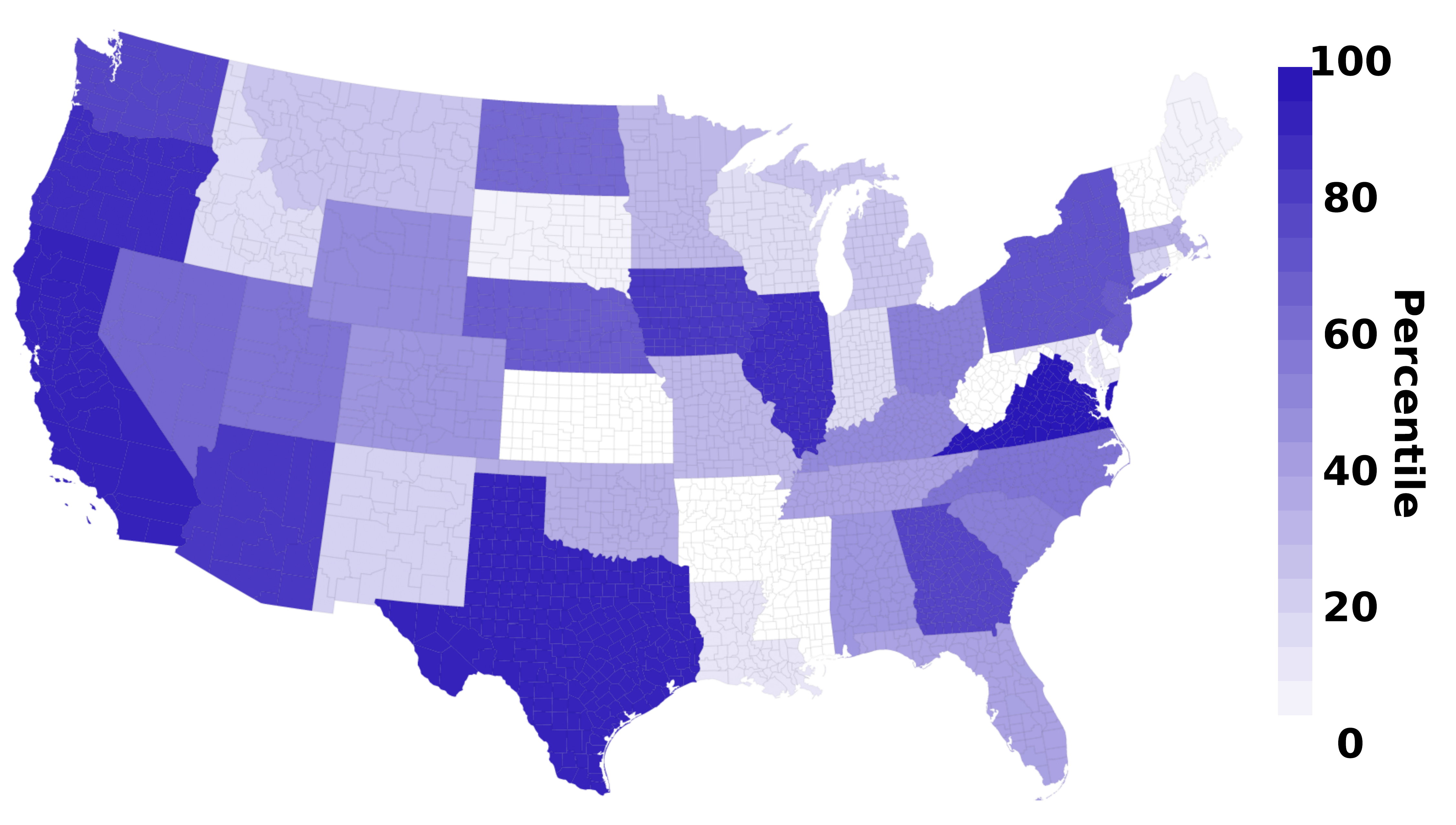}       
    }
    \subfloat[Electricity consumption by state (descending order)]{
        \adjustbox{valign=b}{
        \centering
        \tiny
        \begin{tabular}{c|c|c|c|c|c} 
        \toprule
        \textbf{State} & \begin{tabular}[c]{@{}c@{}}\textbf{Electricity }\\\textbf{Consumption}\\(TWh) \end{tabular} & \textbf{State} & \begin{tabular}[c]{@{}c@{}}\textbf{Electricity }\\\textbf{Consumption}\\(TWh) \end{tabular} & \textbf{State} & \begin{tabular}[c]{@{}c@{}}\textbf{Electricity }\\\textbf{Consumption}\\(TWh) \end{tabular} \\ 
        \hline
        VA & 33.85 & OH & 2.36 & ID & 0.15 \\ 
        \hline
        TX & 21.81 & SC & 2.02 & WI & 0.15 \\ 
        \hline
        CA & 9.33 & WY & 1.86 & MD & 0.10 \\ 
        \hline
        IL & 7.45 & KY & 1.62 & LA & 0.08 \\ 
        \hline
        OR & 6.41 & CO & 1.51 & SD & 0.07 \\ 
        \hline
        AZ & 6.25 & AL & 1.49 & ME & 0.03 \\ 
        \hline
        IA & 6.19 & FL & 1.38 & NH & 0.02 \\ 
        \hline
        GA & 6.18 & TN & 1.33 & RI & 0.02 \\ 
        \hline
        WA & 5.17 & OK & 1.23 & KS &  $<$ 0.01 \\ 
        \hline
        PA & 4.59 & MA & 1.06 & AR & $<$ 0.01 \\ 
        \hline
        NY & 4.07 & MO & 0.97 & DE & $<$ 0.01 \\ 
        \hline
        NJ & 4.04 & MN & 0.82 & DC & $<$ 0.01 \\ 
        \hline
        NE & 3.96 & MT & 0.58 & MS & $<$ 0.01 \\ 
        \hline
        ND & 3.92 & MI & 0.53 & VT & $<$ 0.01 \\ 
        \hline
        NV & 3.42 & NM & 0.40 & WV & $<$ 0.01 \\ 
        \hline
        NC & 2.67 & CT & 0.26 & & \\ 
        \hline
        UT & 2.56 & IN & 0.19 & & \\
        \bottomrule
        \end{tabular}
        }}
        \caption{State-level electricity consumption of U.S. data centers in 2023 \cite{DataCenter_Energy_EPRI_AI_9Percent_US_2030_WhitePaper_2024}.}\label{fig:electricity_consumption_us_datacenter_2023}
\end{figure}

We show in Fig.~\ref{fig:electricity_consumption_us_datacenter_2023} the state-wide data center electricity consumption 
in 2023 \cite{DataCenter_Energy_EPRI_AI_9Percent_US_2030_WhitePaper_2024}.
It can be seen that Virginia, Texas and California have the highest
data center electricity consumption in 2023. The total national electricity consumption reported by EPRI is slightly lower than the values in \cite{DoE_DataCenter_EnergyReport_US_2024}, and we scale it up accordingly in our calculations to ensure consistency.

\section{Additional Results}

\subsection{County-level Public Health Costs of U.S. Data Centers in 2023}\label{appendix:county_level_total_2023}
\begin{figure}[!t]
\centering
    \subfloat[County-level health cost distribution]{
    \includegraphics[height=0.18\linewidth,valign=b]{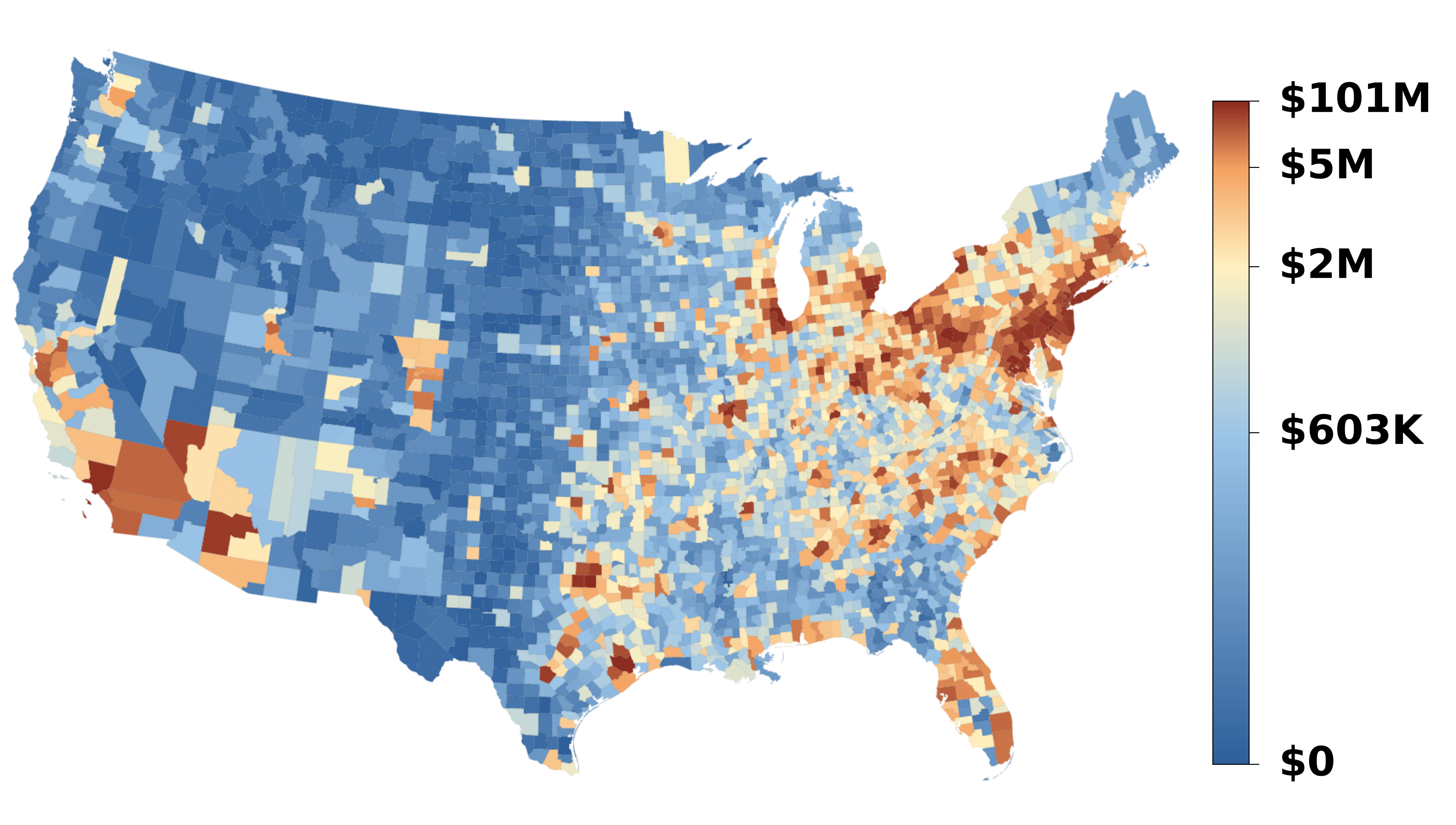}  
    }
    \subfloat[CDF]{
    \includegraphics[height=0.18\linewidth,valign=b]{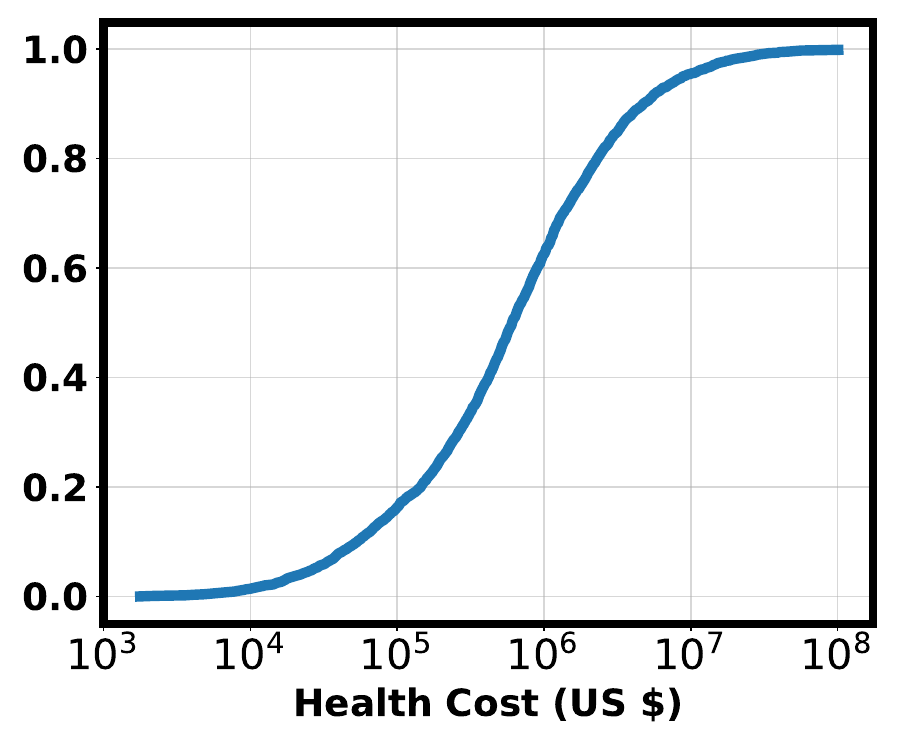}\label{fig:county_level_cdf_2023}         
    }
    \subfloat[Top-10 counties]{
        \adjustbox{valign=b}{
        \centering
        \tiny
        \label{table:top_county_level_cdf_2023}
        \begin{tabular}{c|c|c} 
        \toprule
        \textbf{State} & \textbf{County} & \begin{tabular}[c]{@{}c@{}}
        \textbf{Health Cost} \\ (million \$)\end{tabular} \\ 
        \hline
        IL & Cook & \textbf{101.0} (69.5, 132.4) \\ 
        \hline
        PA & Allegheny & \textbf{88.7} (77.8, 109.6) \\ 
        \hline
        TX & Harris & \textbf{81.5} (56.0, 107.0) \\ 
        \hline
        PA & Philadelphia & \textbf{56.9} (40.2, 73.6) \\ 
        \hline
        OH & Hamilton & \textbf{53.8} (40.8, 66.8) \\ 
        \hline
        MI & Wayne & \textbf{53.5} (37.9, 69.1) \\ 
        \hline
        TX & Dallas & \textbf{50.6} (38.5, 62.6) \\ 
        \hline
        NY & Kings & \textbf{49.5} (31.0, 68.0) \\ 
        \hline
        OH & Cuyahoga & \textbf{48.0} (34.7, 61.3) \\ 
        \hline
        OH & Franklin & \textbf{47.2} (34.8, 59.7) \\ 
        \bottomrule
        \end{tabular}
        }}
        \caption{The county-level total health cost of U.S. data centers in 2023. (a) Health cost map using same percentil-anchored color scale; (b) CDF of county-level health cost;
        (c) Top-10 counties by total health cost.}\label{figure:county_level_total_2019_2023}
\end{figure}

We show in Fig.~\ref{figure:county_level_total_2019_2023}
the county-level total public health cost of U.S. data centers
in 2023, which 
exhibits significant spatial variability. In particular, populated counties located downwind of power plants supplying electricity to data centers tend to experience higher health costs, reflecting the transport of air pollutants across regions. The cumulative distribution function (CDF) in Fig.~\ref{fig:county_level_cdf_2023} highlights that while many counties incur relatively low health costs, a small fraction of counties bear substantially higher impacts. Table~\ref{table:top_county_level_cdf_2023} further identifies the top-10 counties with the highest total health costs, illustrating how local population density and proximity to power generation infrastructure may combine to amplify public health risks in specific counties.

\subsection{Relative to On-road Vehicles}

Section~\ref{sec:relative_fuel_combusion} compares the estimated health impacts of data centers with those of stationary fuel-combustion sectors. Here, we provide a cross-sector comparison focused on on-road emissions (Appendix~\ref{appendix:on_road}). As before, this comparison is intended only to provide a sense of scale. It should not be interpreted as suggesting that a higher or lower estimated impact in one sector affects the importance of mitigating public health impacts in other sectors.
In particular, 
 we consider on-road emissions in California, which has about 35 million registered vehicles and exhibits the highest public health cost from on-road emissions among all U.S. states \cite{California_Vehicle_Registered_35million_2023,Health_COBRA_EPA_Website}.

In 2023, the estimated total public health cost of U.S. data centers is equivalent to 42\% of that from California's on-road emissions. At the national level, the health costs of on-road emissions are estimated to generally decline from 2023 to 2028, reflecting, among other factors, increasingly stringent air-pollutant regulations for vehicles \cite{EPA_Vehicle_LightDuty_Emission_Final_Rule_Summary_Website}. 
Based on the low- and high-growth data center demand scenarios considered in \cite{DoE_DataCenter_EnergyReport_US_2024}, the estimated total public health impact of U.S. data centers could reach \$11.7 billion and \$20.9 billion in 2028, respectively. Under the high-growth scenario, this estimated health burden could approach the scale of California's on-road emissions. This comparison is intended to illustrate that the estimated health externalities associated with U.S. data centers may grow relatively faster under the scenarios considered, rather than to suggest that the data centers and on-road vehicles are directly comparable or that their mitigation strategies are interchangeable.

\section{Health-Informed Computing}

Addressing air-pollution-related health impacts requires coordinated efforts across sectors, along with mitigation strategies tailored to each sector \cite{Health_COBRA_EPA_Website}. While air pollution health impacts from sectors such as transportation have been widely studied \cite{Health_ElectricVehicle_PowerPlant_Toronto_PNAS_2024_schmitt2024health}, the public health impacts of data centers have received comparatively less attention.
Here, we present Health-Informed Computing, a framework that explicitly incorporates public health impacts as a key optimization objective and strategically manages data center workloads to minimize adverse health outcomes while supporting broader sustainability goals.

To mitigate the public health impact of data centers, one straightforward approach is to focus solely on reducing the energy consumption. While reducing energy consumption is beneficial, overlooking the downstream public health impact of \emph{where} and \emph{when} energy is produced does not necessarily lead to minimized health burdens. For example, Table~\ref{table:health_cost_llama31}
demonstrates a 10x difference in health costs for training
the same AI model across different locations.  This suggests that health-informed and energy-aware computing, when combined, can offer complementary benefits, leading to better public health outcomes.

\subsection{Opportunities for Health-Informed Computing}\label{appendix:health_informed_opportunity}

\begin{figure}[!t]
\centering
    \subfloat[Normalized IQR (110)]{\includegraphics[width=0.3\textwidth,valign=b]{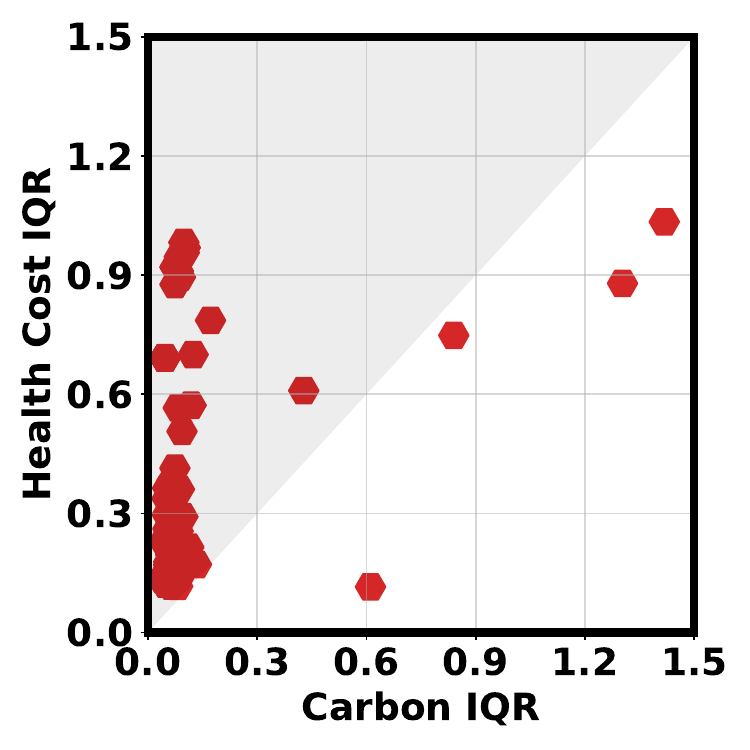}\label{carbon_health_IQR_2024_year}    
    }
    \subfloat[Normalized STD (90)]{\includegraphics[width=0.3\textwidth,valign=b]{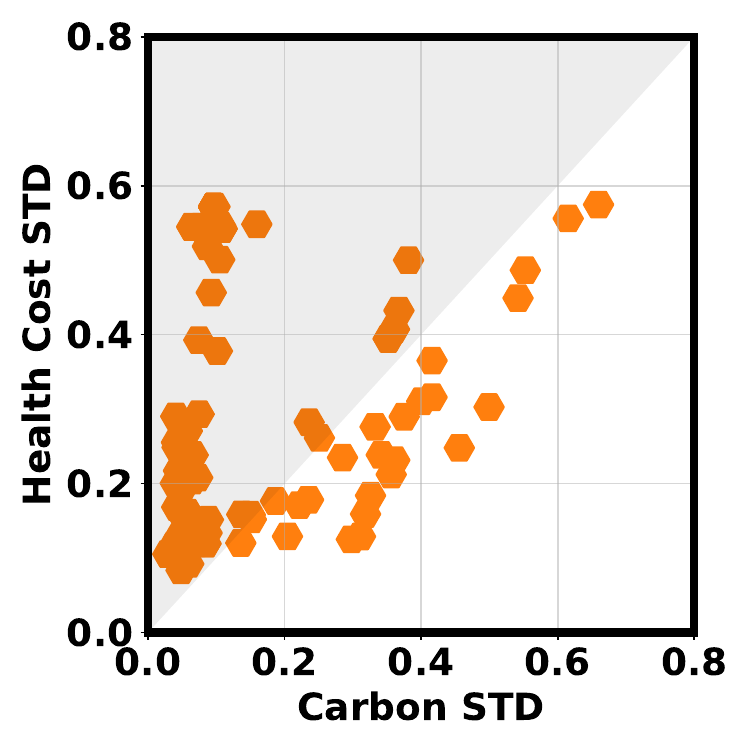}\label{carbon_health_STD_2024_year}          
    }
    \subfloat[Average]{\includegraphics[width=0.3\textwidth,valign=b]{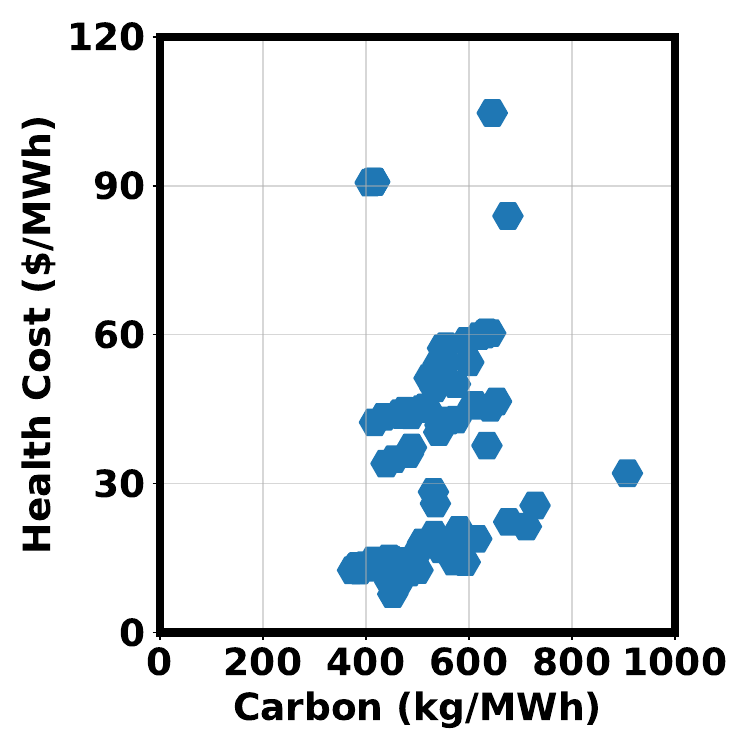}\label{carbon_health_avg_2024_year}          
    }
    \caption{Analysis of marginal scope-2 carbon emission rates and public health costs
    over 114 U.S. regions between October 1, 2023 and September 30, 2024 \cite{WattTime_Website}. 
    (a) In 110 out of the 114 U.S. regions (96\%), the normalized IQR of marginal health cost is higher
    than that of marginal carbon intensity. (b) In 90 out of the 114 U.S. regions (79\%), the normalized standard deviation of marginal health cost is higher than that of marginal carbon intensity.
    (c) The Pearson correlation between
    the per-region yearly average marginal health cost and carbon intensity is 0.292.}\label{fig:carbon_health_analysis}
\end{figure}

Data centers, including those operated
by major technology companies \cite{Google_SustainabilityReport_2024,Facebook_SustainabilityReport_2024}, 
mostly rely on grid electricity due to the practical challenges of installing on-site low-pollutant and low-carbon energy sources at scale. 
However, the spatial-temporal variations of scope-2 health prices  (Fig.~\ref{fig:carbon_health_analysis}) open up new opportunities
to reduce the public health impact by exploiting the high 
scheduling flexibilities of computing workloads
(e.g., AI training).   
For example, as further supported by EPRI's recent initiative
on maximizing data center flexibility for demand response \cite{DataCenter_FlexibilityInitiative_EPRI_WhitePaper_2024}, 
AI training can be scheduled
in more than one data center, while multiple AI models with different
sizes are often available to serve AI inference requests,
offering flexible resource-performance tradeoffs.

To date,  the existing data centers
have mostly exploited 
such scheduling flexibilities for reducing electricity costs \cite{DataCenter_CuttingElectricBill_MIT_Sigcomm_2009_10.1145/1592568.1592584},
carbon emissions \cite{Google_CarbonAwareComputing_PowerSystems_2023_9770383}, water consumption \cite{Shaolei_Water_SpatioTemporal_GLB_TCC_2018_7420641},
and/or environmental inequity \cite{Shaolei_Equity_GLB_Environmental_AI_eEnergy_2024}.
Nonetheless, the public health impact can differ from these environmental costs or metrics.

Concretely, despite sharing some common sources 
(e.g., fossil fuels) with carbon emissions, 
the public health impact resulting from the dispersion of criteria air pollutants is highly dependent on the emission source location and only exhibits
a weak correlation with carbon emissions. For example, the same quantity of carbon emissions generally results in the same climate change impacts regardless of the emission source; in contrast, criteria air pollutants have substantially greater public health impacts if emitted in densely populated regions compared to sparsely populated or unpopulated regions, emphasizing the importance of considering spatial variability.

To further confirm this point and highlight the potential
of health-informed data center load shifting, we analyze
the scope-2 marginal carbon intensity and public health cost for each unit of electricity generation across all the 114 U.S. regions between October 1, 2023, and September 30, 2024, provided by WattTime \cite{WattTime_Website}.\footnote{The health cost signal provided by \cite{WattTime_Website} only considers
mortality from \pmtwo, while COBRA includes a variety of health outcomes including asthma, lung cancer, and mortality from ozone, among others \cite{Health_COBRA_EPA_Mannual}.}  The time granularity for data collection is 5 minutes.

Here, we focus on marginal health impacts and carbon emissions for two main reasons: first, WattTime provides only real-time marginal health impact estimates \cite{WattTime_HealthDamage_Intro_Website};
and second, marginal signals are sometimes considered more useful for guiding energy load adjustments \cite{Carbon_Marginal_Average_DataCenter_Shifting_LineRoald_gorka2024electricityemissionsjlframeworkcomparisoncarbon}, which also  explain why the EPA reports marginal health benefits per kWh (i.e., marginal health price) to inform energy demand changes \cite{Health_Benefit_kWh_EPA_Website}.

We show in Fig.~\ref{carbon_health_IQR_2024_year} the region-wise normalized interquartile ranges (IQR divided
by the yearly average) for both public health costs and carbon emissions.
The normalized IQR measures the spread of the time-varying health
and carbon signals. Specifically,  in 110 out of the 114 U.S. regions (96\%), the normalized IQR of health cost is higher
    than that of the carbon intensity for each unit of electricity consumption. Moreover,
the normalized IQR for carbon emissions is less than 0.2 in most of the regions. This implies that health costs
    exhibit a greater temporal variation than carbon emissions
    in 110 out of the 114 U.S. regions. 
Likewise, in Fig.~\ref{carbon_health_STD_2024_year}, the greater temporal variation of health
costs is also supported by its greater normalized standard deviation
(STD divided by the yearly average) 
in 90 out of the 114 U.S. regions (79\%).
Next, we show in Fig.~\ref{carbon_health_avg_2024_year} the weak spatial correlation (Pearson correlation coefficient: 0.292)
between the yearly average health cost and carbon intensity across the 114 regions. Furthermore, the normalized IQR 
of the health cost spatial distribution is 3.62x that
of carbon emission spatial distribution (1.05 vs. 0.29),
while the health-to-carbon ratio in terms of the spatial distribution's normalized STD 
 is 3.37 (0.64 vs. 0.19).
In other words, the health cost could have a greater spatial spread
than the carbon emission.

\begin{table}[!tp]
\scriptsize
\centering
\begin{tabular}{l|c|c|c|c|c|c|c} 
\toprule
\multirow{2}{*}{\textbf{Location}} & \multirow{2}{*}{\begin{tabular}[c]{@{}c@{}} \textbf{Pearson}\\\textbf{Correlation}\end{tabular}} & \multicolumn{3}{c|}{\textbf{Normalized IQR} } & \multicolumn{3}{c}{\textbf{Normalized STD} } \\ 
\cline{3-8}
 & & Health & Carbon & $\frac{\text{Health}}{\text{Carbon}}$ Ratio & Health & Carbon & $\frac{\text{Health}}{\text{Carbon}}$ Ratio \\ 
\hline
Loudoun County, VA & 0.427 & 0.158 & 0.065 & 2.409 & 0.131 & 0.059 & 2.222 \\ 
\hline
Central Ohio, OH & 0.479 & 0.160 & 0.065 & 2.441 & 0.137 & 0.066 & 2.064 \\ 
\hline
The Dalles, OR & 0.326 & 0.957 & 0.099 & 9.614 & 0.546 & 0.103 & 5.296 \\ 
\hline
Douglas County, GA & 0.756 & 0.507 & 0.093 & 5.418 & 0.293 & 0.075 & 3.913 \\ 
\hline
Montgomery County, TN & 0.760 & 0.289 & 0.067 & 4.320 & 0.195 & 0.046 & 4.236 \\ 
\hline
Papillion, NE & 0.736 & 0.748 & 0.840 & 0.891 & 0.487 & 0.553 & 0.881 \\ 
\hline
Storey County, NV & 0.584 & 0.178 & 0.057 & 3.132 & 0.168 & 0.042 & 4.004 \\ 
\hline
Ellis County, TX & 0.474 & 0.196 & 0.082 & 2.384 & 0.232 & 0.361 & 0.641 \\ 
\hline
Berkeley County, SC & 0.416 & 0.156 & 0.054 & 2.911 & 0.105 & 0.044 & 2.405 \\ 
\hline
Council Bluffs, IA & 0.361 & 0.185 & 0.111 & 1.671 & 0.129 & 0.311 & 0.415 \\ 
\hline
Henderson, NV & 0.584 & 0.178 & 0.057 & 3.132 & 0.168 & 0.042 & 4.004 \\ 
\hline
Jackson County, AL & 0.760 & 0.289 & 0.067 & 4.320 & 0.195 & 0.046 & 4.236 \\ 
\hline
Lenoir, NC & 0.240 & 0.176 & 0.059 & 2.982 & 0.129 & 0.046 & 2.800 \\ 
\hline
Mayes County, OK & 0.617 & 0.122 & 0.049 & 2.495 & 0.171 & 0.222 & 0.772 \\
\bottomrule
\end{tabular}
\caption{Correlation analysis of marginal carbon emissions and health impacts for a technology company's U.S. data center locations between October 1, 2023, and September 30, 2024 \cite{WattTime_Website}. 
According to the region classification of WattTime \cite{WattTime_HealthDamage_Intro_Website}, the two data centers in Storey County, NV, and Henderson, NV, belong
to the same power grid region, and so do those in Jackson County, AL, and Montgomery County, TN.}\label{table:correlation_google_datacenter}
\end{table}
In addition to analysis for all U.S. regions,
 we  turn to specific regions where
a large technology company builds its U.S. data centers. 
 We present the results Table~\ref{table:correlation_google_datacenter},
further confirming
 that carbon intensities and health impacts are not always aligned and that health impacts vary more significantly
than carbon intensities in almost all the locations.

We further analyze the Pearson correlation coefficients between
hourly marginal health prices and carbon emission rates throughout 2023 for the U.S. regions that have complete health
and carbon data provided by WattTime \cite{WattTime_Website}. 
Nearly 70\% of
the regions have a weak or moderate correlation,
with a carbon-health correlation coefficient of less
than 0.60. This implies that despite having
fossil fuels as the common source, health costs
and carbon emissions are different and can exhibit trade-offs. Moreover, due to their additional dependence on population distribution
and meteorological conditions, health prices often demonstrate more pronounced temporal fluctuations than carbon emissions.

These findings suggest that leveraging spatiotemporal variations in a health-aware manner can reduce the public health costs of data center operations. 
Moreover, the observed distinctions between health impacts and carbon emissions suggest the need to optimize data center decisions by explicitly accounting for and exploiting the spatiotemporal heterogeneity of health impacts.

\subsubsection{Location-Dependent Public Health Impacts of Two Technology Companies}

We further highlight 
 locational dependency of public health impacts by considering
 two major technology companies' U.S. data center locations in 2023, excluding
their leased colocation data centers whose locations are proprietary.
We name these two companies A and B, respectively. These two companies do not have same data center locations.
While company B discloses its per-location electricity usage \cite{Facebook_SustainabilityReport_2024}, company A does not.
Thus,
we uniformly distribute company A's North America electricity
consumption over its U.S. data center locations based on its latest
sustainability report \cite{Google_SustainabilityReport_2024}.
We consider location-based emission accounting
without taking into account renewable energy credits these two companies apply
to offset their grid electricity consumption.
Note that our results are intended to illustrate the locational dependence of public health impacts and should not be interpreted as a precise assessment of the actual impacts of these two companies.

\begin{figure}[!tp]
    \centering
    \subfloat[Per-household health cost (Company A)]{
   \includegraphics[width=0.45\textwidth,valign=b]{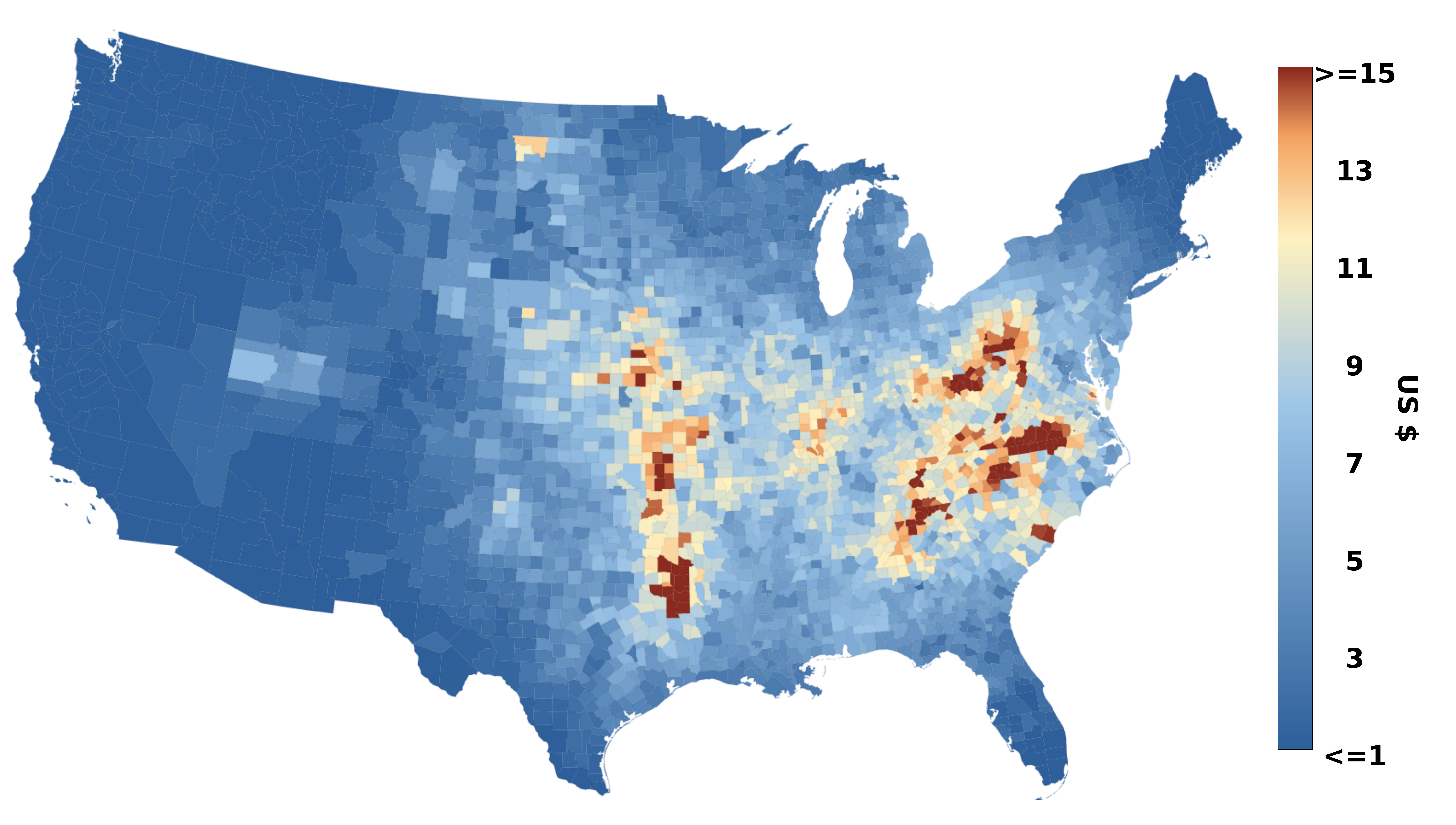} 
    }
    \subfloat[Per-household health cost (Company B)]{
    \includegraphics[width=0.45\textwidth,valign=b]{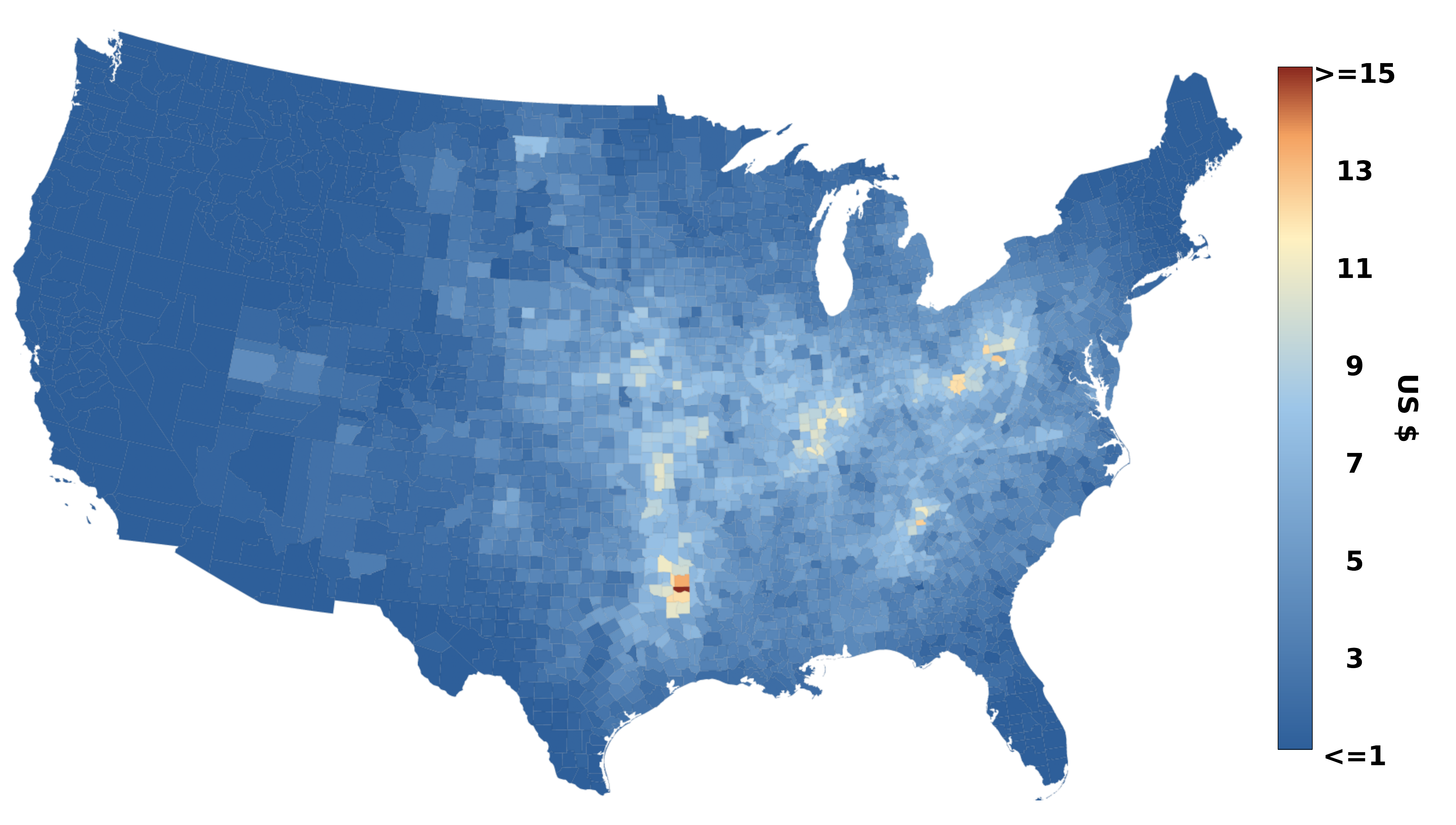}        
    }
    \caption{The county-level per-household health cost of two U.S. technology companies in 2023. The legend uses a linear value scale from \$1 to \$15 per household, with values below \$1 and above \$15 shown at the endpoints.}\label{fig:perhousehold_tech_company_average_2023}
\end{figure}

We see from Fig.~\ref{fig:perhousehold_tech_company_average_2023} 
that the two companies have different per-household health cost
distributions and most-affected counties. This is partly due
to the two companies' different data center locations, and highlights
the locational dependency of public health impacts. 
That is, unlike carbon emissions that have a similar effect regardless of the
emission source locations, the public health impact of criteria air pollutants heavily depends on the location of the
emission source. Thus, technology companies should account for
public health impacts when
deciding where they build data centers,
where they get electricity for their data centers, and where they install  renewables in order to best mitigate public health impacts. 

\begin{table}[!ht]
\centering
\scriptsize
\caption{The public health cost of training a large AI model 
in selected U.S. data centers.}
\label{table:health_cost_llama31}
\begin{tabular}{c|c|c|c|c|c|c|c} 
\hline
\multirow{2}{*}{\textbf{Location}} & 
\multirow{2}{*}{\begin{tabular}[c]{@{}c@{}}\textbf{Electricity Price} \\(\textcent/kWh)\end{tabular}} & 
\multirow{2}{*}{\begin{tabular}[c]{@{}c@{}}\textbf{Electricity} \\(million \$)\end{tabular}} & 
\multirow{2}{*}{\begin{tabular}[c]{@{}c@{}}\textbf{Health Cost}\\(million \$)\end{tabular}} & 
\multirow{2}{*}{\begin{tabular}[c]{@{}c@{}}\textbf{\% of Electricity} \\
\textbf{Cost}\end{tabular}} &
\multicolumn{3}{c}{\textbf{Emission} (Metric Ton)} \\
\cline{6-8}
& & & & & \textbf{PM2.5} (LA-NYC) & \textbf{NOx} (LA-NYC) & \textbf{SO2} \\
\hline
Huntsville, AL&7.11 &2.1 &\textbf{0.70} (0.54, 0.87) &33\%&0.61 (13800)&2.80 (2500)&2.72\\
\hline
Stanton Springs, GA& 6.88&2.0&\textbf{0.85} (0.65, 1.04)&41\%&0.69 (15500)&3.37 (3000)&3.35\\
\hline
DeKalb, IL&8.20 &2.4 &\textbf{1.92} (1.41, 2.42)&79\%&1.25 (28100)&7.31 (6600)&7.83\\
\hline
Altoona, IA& 6.91&2.1 &\textbf{2.51} (1.84, 3.17)&122\%&1.52 (34000)&11.78 (10600)&14.76\\
\hline
Sarpy, NE& 7.63&2.3 & \textbf{1.54} (1.16, 1.92)&68\%&1.13 (25300)&13.5 (12200)&18.51\\
\hline
Los Lunas, NM& 5.75&1.7 &\textbf{0.73} (0.56, 0.90) &43\%&0.78 (17500)&8.36 (7500)&9.84\\
\hline
Forest City, NC& 7.15&2.1 &\textbf{1.07} (0.85, 1.30) &50\%&0.72 (16200)&5.72 (5200)&3.27\\
\hline
New Albany, OH& 7.03&2.1 &\textbf{1.61} (1.20, 2.03) &77\%&1.13 (25200)&5.15 (4600)&4.44\\
\hline
Prineville, OR& 7.52&2.2 &\textbf{0.23} (0.19, 0.28) &10\%&0.59 (13300)&4.67 (4200)&2.40\\
\hline
Gallatin, TN& 6.23&1.9 &\textbf{0.32} (0.24, 0.40) &17\%&0.41 (9200)&1.21 (1100)&0.93\\
\hline
Fort Worth, TX& 6.60&2.0 &\textbf{0.51} (0.38, 0.65)&26\%&0.47 (10500)&3.02 (2700)&3.81\\
\hline
Eagle Mountain, UT& 6.99&2.1&\textbf{0.24} (0.19, 0.29) &12\%&0.60 (13300)&4.82 (4300)&2.52\\
\hline
Henrico, VA&8.92 &2.7 &\textbf{1.61} (1.20, 2.03) &61\%&1.13 (25200)&5.15 (4600)&4.44\\
\hline
\end{tabular}
\end{table}

\subsubsection{Location-Dependent Public Health Impact of Generative AI Training}

We next examine the estimated health impact of a specific computing task to illustrate how the same workload can result in different public health costs depending on where it is performed. Specifically, because AI training workloads often have substantial spatial flexibility,
we consider the training of an LLM and assume electricity consumption comparable to that reported for training Meta's Llama-3.1 \cite{Llama_31_Introduction_Meta_Website_2024}. Because scope-2 impacts dominate in our analysis, and because the power allocation and on-site backup generation usage for training Llama-3.1 are not publicly known, we focus on scope-2 health costs associated with electricity consumption.

Importantly, although we use the estimated electricity consumption of Llama-3.1 and a set of U.S. data center locations as an illustrative example, our results should be interpreted as estimates for training a general LLM at a scale comparable to Llama-3.1, rather than as an assessment of the actual health impact of Llama-3.1. The actual health impacts of Llama-3.1 depend on specific factors such as timing, electricity procurement, grid mix, and any on-site generation use, and may differ substantially from our estimates.

We show the results in Table~\ref{table:health_cost_llama31}. It
can be seen that the total health cost 
varies widely depending on the training data center locations. For example,
the total health cost is only \$0.23 million in Oregon, whereas
the cost will increase dramatically to \$2.5 million in Iowa due to various factors, such as the wind direction and the 
pollutant emission rate for electricity generation \cite{Health_AVERT_Emission_Marginal_EPA_Website}.

Additionally, depending on the data center location, training an AI model at the scale of Llama-3.1 can be associated with air-pollutant emissions
comparable to those from more than 10,000 LA-NYC round trips by car. This cross-sector comparison is intended only to provide a sense of emissions scale. It should not be interpreted as discouraging either activity or as suggesting that the two activities have the same public health impacts. The health impacts depend strongly on factors such as pollutant mix, emissions location, atmospheric dispersion, and exposed population density.
Quantitatively, our estimates suggest that in some locations, the public health externalities of AI model training can be comparable to the associated electricity costs.

Overall, the results highlight that the public health impact of AI model training is highly location-dependent. Combined with the spatial flexibility of model training, they suggest that
AI model developers can take into account
potential health impacts when choosing data center locations for training where applicable.

\subsection{Formulation of Health-Informed Computing}\label{sec:health_informed_benefit}

To improve system performance and reliability, technology companies typically operate data centers over a variety of geographically distributed regions and \emph{dynamically} distribute computing workloads through a process known as geographical load balancing (GLB). 
Here, we leverage the unique spatial load flexibility of geographically
distributed data centers to demonstrate
the benefits of health-informed computing as a proof of concept.

Specifically, we study {health-informed} GLB as an example of \ouralg to mitigate the public health impacts of data center operation.
We consider a discrete-time model 
of duration $T$
and measure the workloads in terms of their energy demand. For the sake of examining the impact
of spatial flexibility, we assume that
the energy loads (i.e., workloads) can be flexibly distributed across a set of $N$ data centers denoted as $\mathcal{N} = \{1, 2, \ldots, N\}$. In each time $t \in \{1, 2, \ldots, T\}$, the total energy demand is $M_t$, and $w_{i,t}$ represents the load assigned to data center $i$. We use $l_i$ to represent the default load capacity of data center $i$, and introduce a slackness parameter $\lambda\geq1$ to represent
the ability of each data center to accept loads in excess of its default capacity.
Thus, the greater the value of $\lambda$,
the more spatial flexibility the operator has.
In each time $t$, we use $p_{i, t}^e$ and $p_{i, t}^h$ to denote the electricity price and health price at data center $i$, respectively.

The health price $p_{i, t}^h$ depends on the emission source of criteria air pollutants (e.g., power plants' emission rates and their locations),
air pollutant dispersion, and estimates of adverse health outcomes and resulting costs attributed to the increased air pollutant concentration in each region \cite{Health_COBRA_EPA_Mannual}. Thus, the health price quantifies
the ultimate health burden  
imposed on affected populations and is measured
in terms of economic costs
for each unit of electricity consumption. 
It varies over time due to fluctuations in the grid’s generation mix and changing meteorological conditions. Third-party organizations such as WattTime \cite{WattTime_HealthDamage_Intro_Website} provide real-time estimates of the marginal health price of electricity across 114 power balancing regions in the U.S., while the EPA \cite{Health_Benefit_kWh_EPA_Website} reports annualized average health prices for electricity in 14 broader regions nationwide.

A canonical objective in GLB is carbon-aware computing, which dynamically
dispatches workloads to data centers with lower carbon intensities.
To further incorporate public health impacts, the GLB objective for each time slot
$t$ can be written as:
\begin{subequations}
\begin{align}
        &\min_{\mathbf{w}_{t}\in\mathcal{W}_t}\sum_{i = 1}^N \left(p_{i,t}^e + p_{i,t}^h + p^c r_{i,t}^c\right)\cdot w_{i,t}, \\
        \mathrm{s.t.} & \qquad \sum_{i=1}^N w_{i, t} = M_t,
        \label{eqn:total_workload_cons}\\
         &\qquad 0 \le w_{i, t} \le \lambda \cdot l_i,\quad \forall i \in \mathcal{N}, \label{eqn:upper_limit_cons}
    \end{align}
\end{subequations}
where $N$ is the number of data centers,
$p_{i,t}^e$, $p_{i,t}^h$, and $r_{i,t}^c$ denote, respectively,
the electricity price, health cost, and carbon emissions per kWh at
data center $i$ and time $t$,
$p^c$ denotes the carbon price,
$\mathbf{w}_{t}=(w_{1,t},\cdots,w_{N,t})$ denotes the energy use induced by
the dispatched workloads, 
the constraint \eqref{eqn:total_workload_cons} means that
all loads must be dispatched to a data center (with no loads dropped),  the constraint \eqref{eqn:upper_limit_cons} encodes the maximum workload capacity of each 
data center parameterized by the slackness factor $\lambda\geq1$,
and
$\mathcal{W}_t$ denotes the feasible set at time $t$, capturing operational
constraints such as latency constraints and workload-specific requirements \cite{Gao:2012:EG:2377677.2377719}.
 Carbon prices commonly fall in the range of \$10 to \$200 per ton, depending on the market and compliance requirements

Our formulation can be easily
extended to incorporate additional considerations. For example, it can include
long-term, per-region health impact constraints, rather than focusing solely on national-level total health costs.
For the sake of clarity, we set these extensions aside to focus on the novel metric of health cost for data center resource management.

\subsection{Sensitivity Analysis of Capacity Slackness $\lambda$}

The parameter $\lambda \geq 1$ represents the amount of capacity slackness for accommodating additional workloads: a larger $\lambda$ provides greater spatial flexibility. We vary $\lambda$ and report the results for $\lambda=1.2$ and $\lambda=2.0$ in Tables~\ref{table_results_glb_12} and~\ref{table_results_glb_20}, respectively.  As $\lambda$ increases, the additional spatial flexibility leads to larger health benefits. In particular, when $\lambda=2.0$, the health-oblivious GLB algorithm ($p_{i,t}^{h}=0$) reduces carbon emissions by nearly 10\%, but lowers health cost by less than 5\% relative to the baseline. By contrast, \hico\ incorporates health into the scheduling objective. Although carbon emissions and health cost cannot in general be minimized at the same time, a suitable choice of $p^c$ allows \hico\ to substantially reduce health cost while also achieving meaningful reductions in carbon emissions and electricity cost. For example, under $\lambda=2.0$ and $p^c=\$200/\text{ton}$, \hico\ reduces carbon emissions by nearly 7\% while lowering health cost by more than 30\%. More broadly, these results suggest that health-informed scheduling can enable data center operation to jointly improve public health, environmental sustainability, and economic efficiency.

\begin{table*}[!t]
\centering
\caption{\ouralg under different settings ($\lambda=1.2$), with percent changes
in parentheses reported relative to the baseline.
Three special cases:
Electricity Cost Optimal (\textbf{\eopt}, with $p^c=0$ and $p^h_{i,t}=0$),
Carbon Emission Optimal (\textbf{\copt}, with $p^e_{i,t}=0$ and $p^h_{i,t}=0$),
and Health Impact Optimal (\textbf{\hopt}, with $p^e_{i,t}=0$ and $p^c=0$).
}
\vspace{-0.3cm}
\label{table_results_glb_12}
\scriptsize
\setlength{\tabcolsep}{3pt}
\renewcommand{\arraystretch}{1.15}
\resizebox{\textwidth}{!}{%
\begin{tabular}{c|c|cc|cc|cc|cc|cc|cc|cc|cc}
\toprule
\multirow{2}{*}{\textbf{Metric}} &
\multirow{2}{*}{Baseline} &
\multicolumn{2}{c|}{\multirow{2}{*}{\eopt}} &
\multicolumn{2}{c|}{\multirow{2}{*}{\copt}} &
\multicolumn{2}{c|}{\multirow{2}{*}{\hopt}} &
\multicolumn{4}{c|}{Health-Oblivious ($p^h_{i,t}=0$)} &
\multicolumn{6}{c}{Health-Informed ($p^h_{i,t}\not=0$)} \\
\cline{9-18}
&& \multicolumn{2}{c|}{} &
\multicolumn{2}{c|}{} &
\multicolumn{2}{c|}{} &
\multicolumn{2}{c|}{$p^c=$\$50/ton} &
\multicolumn{2}{c|}{$p^c=$\$200/ton} &
\multicolumn{2}{c|}{$p^c=$\$0/ton} &
\multicolumn{2}{c|}{$p^c=$\$50/ton} &
\multicolumn{2}{c}{$p^c=$\$200/ton} \\
\hline
Health (Million \$) &
393.23 &
374.20 & (-4.84\%) &
404.72 & (+2.92\%) &
\textbf{345.26} & \textbf{(-12.20\%)} &
377.01 & (-4.13\%) &
395.28 & (+0.52\%) &
349.29 & (-11.17\%) &
349.10 & (-11.22\%) &
350.34 & (-10.91\%) \\
\hline
Energy (Million \$) &
756.50 &
\textbf{732.06} & \textbf{(-3.23\%)} &
761.20 & (+0.62\%) &
752.26 & (-0.56\%) &
733.25 & (-3.07\%) &
747.88 & (-1.14\%) &
738.49 & (-2.38\%) &
739.32 & (-2.27\%) &
744.67 & (-1.56\%) \\
\hline
Carbon (Million Ton) &
6.60 &
6.68 & (+1.22\%) &
\textbf{6.37} & \textbf{(-3.54\%)} &
6.54 & (-0.87\%) &
6.53 & (-1.14\%) &
6.40 & (-3.04\%) &
6.57 & (-0.44\%) &
6.54 & (-0.89\%) &
6.49 & (-1.71\%) \\
\bottomrule
\end{tabular}
}
\end{table*}

\begin{table*}[!t]
\centering
\caption{\ouralg under different settings ($\lambda=2.0$), with percent changes
in parentheses reported relative to the baseline.
Three special cases:
Electricity Cost Optimal (\textbf{\eopt}, with $p^c=0$ and $p^h_{i,t}=0$),
Carbon Emission Optimal (\textbf{\copt}, with $p^e_{i,t}=0$ and $p^h_{i,t}=0$),
and Health Impact Optimal (\textbf{\hopt}, with $p^e_{i,t}=0$ and $p^c=0$).
}
\vspace{-0.3cm}
\label{table_results_glb_20}
\scriptsize
\setlength{\tabcolsep}{3pt}
\renewcommand{\arraystretch}{1.15}
\resizebox{\textwidth}{!}{%
\begin{tabular}{c|c|cc|cc|cc|cc|cc|cc|cc|cc}
\toprule
\multirow{2}{*}{\textbf{Metric}} &
\multirow{2}{*}{Baseline} &
\multicolumn{2}{c|}{\multirow{2}{*}{\eopt}} &
\multicolumn{2}{c|}{\multirow{2}{*}{\copt}} &
\multicolumn{2}{c|}{\multirow{2}{*}{\hopt}} &
\multicolumn{4}{c|}{Health-Oblivious ($p^h_{i,t}=0$)} &
\multicolumn{6}{c}{Health-Informed ($p^h_{i,t}\not=0$)} \\
\cline{9-18}
&& \multicolumn{2}{c|}{} &
\multicolumn{2}{c|}{} &
\multicolumn{2}{c|}{} &
\multicolumn{2}{c|}{$p^c=$\$50/ton} &
\multicolumn{2}{c|}{$p^c=$\$200/ton} &
\multicolumn{2}{c|}{$p^c=$\$0/ton} &
\multicolumn{2}{c|}{$p^c=$\$50/ton} &
\multicolumn{2}{c}{$p^c=$\$200/ton} \\
\hline
Health (Million \$) &
393.23 &
368.10 & (-6.39\%) &
408.92 & (+3.99\%) &
\textbf{221.38} & \textbf{(-43.70\%)} &
380.09 & (-3.34\%) &
374.82 & (-4.68\%) &
223.96 & (-43.05\%) &
227.73 & (-42.09\%) &
261.56 & (-33.48\%) \\
\hline
Energy (Million \$) &
756.50 &
\textbf{702.16} & \textbf{(-7.18\%)} &
767.59 & (+1.47\%) &
734.29 & (-2.94\%) &
709.77 & (-6.18\%) &
723.42 & (-4.37\%) &
728.19 & (-3.74\%) &
727.07 & (-3.89\%) &
725.11 & (-4.15\%) \\
\hline
Carbon (Million Ton) &
6.60 &
6.72 & (+1.73\%) &
\textbf{5.84} & \textbf{(-11.49\%)} &
6.61 & (+0.19\%) &
6.06 & (-8.13\%) &
5.95 & (-9.91\%) &
6.55 & (-0.71\%) &
6.43 & (-2.60\%) &
6.16 & (-6.71\%) \\
\bottomrule
\end{tabular}
}
\end{table*}

\end{document}